\RequirePackage{fix-cm}
\documentclass[twocolumn]{svjour3}          
\smartqed  
\usepackage[T1]{fontenc}
\usepackage{graphicx}
\usepackage{float}
\usepackage[numbers]{natbib}
\usepackage{booktabs}
\usepackage{multirow}
\usepackage{graphicx}
\usepackage{caption}
\usepackage{diagbox}
\usepackage{pbox}
\usepackage{placeins}
\usepackage{multicol} 
\usepackage{tikz}
\usetikzlibrary{positioning}
\usepackage{tabulary}
\usepackage{tabularx}
\usepackage{multirow}  
\usepackage{amsmath}

\begin{document}

\title{Toward an Abstract Model of  Programmable Data Plane Devices}



\author{Debobroto Das Robin        \and
        Dr. Javed I. Khan 
}


\institute{F. Author \at
              Kent State University, Kent, Ohio, USA\\
              \email{drobin@kent.edu}           
           \and
           S. Author \at
              Kent State University, Kent, Ohio, USA\\
              \email{javed@kent.edu}
}

\date{Received: date / Accepted: date}

\maketitle

\begin{abstract}
SDN divides the networking landscape into 2 parts: control and data  plane. SDN expanded it's foot mark starting with OpenFlow based highly flexible control plane and rigid data plane. Innovation and improvement in hardware design and development is bringing various new architectures for data plane. Data plane is becoming more programmable then ever before. A common abstract model of data plane is required to develop complex application over these heterogeneous  data plane devices. It can also provide insight about performance optimization and bench-marking of programmable data plane devices. Moreover, to understand and utilize  data plane's programmability, a detailed structural analysis and an identifiable matrix to compare different devices are required. In this work, an improved and structured abstract model of the programmable data plane devices is presented and  features of its components are discussed in detail. Several commercially available programmable data plane devices are also compared based on those features. 
\keywords{ Programmable data plane \and Abstract model   \and Hardware abstraction layer \and P4 \and Software Defined Network (SDN) }
\end{abstract}

\section{Introduction} \label{Introduction} 
 From the very beginning, behavior of data plane devices were rigid and ruled by TCP/IP protocol stack. Different efforts \cite{Feamster:2014:RSI:2602204.2602219} for making this rigid environment more programmable haven't gain momentum until  OpenFlow protocol \cite{McKeown:2008:OEI:1355734.1355746} came to the theater. 
 But, with it's ever growing  protocol field set, OpenFlow can not achieve the goal of real programmable network \cite{Bosshart:2014:PPP:2656877.2656890}. For true network programmability, data plane hardware architecture should be decoupled from protocol and  programmable in nature. Various technologies \cite{openFlowSwitchSpecV135,ODP,hoiland2018express,dpdk,VPP,choi2018fboss,SAI,Bosshart:2013:FMF:2486001.2486011,Chole:2017:DDP:3098822.3098823} have been developed to program the run time behavior of a data plane device \textit{`on the fly'}. They  are  becoming more computationally capable and various complex application layer processing tasks are being pushed to data plane \cite{Sivaraman:2017:HDE:3050220.3063772,Dang:2015:NCN:2774993.2774999,Kohler:2018:PTI:3229591.3229593}. This is enabling emergence of a new paradigm \textit{`in-network computation'} \cite{Sapio:2017:ICD:3152434.3152461,Benson:2019:ICC:3317550.3321436}.  Enabling such cross layer behavior and protocol independence in data plane makes OSI layer based definition of switch  (L2, L3, L4 switch etc.) obsolete. This raises the necessity of structured discussion on: how to define programmable switch/data plane device and how to define it's programmability features? (\textit{Through  rest of the paper, programmable switch and programmable  data plane  (\textbf{\textit{PDP}}) device, these two terms are used interchangeably.}) 

\begin{figure}
	\centering
	\includegraphics[ width=\columnwidth, trim={0.0in 4.55in 6.72in 0.0in},clip]{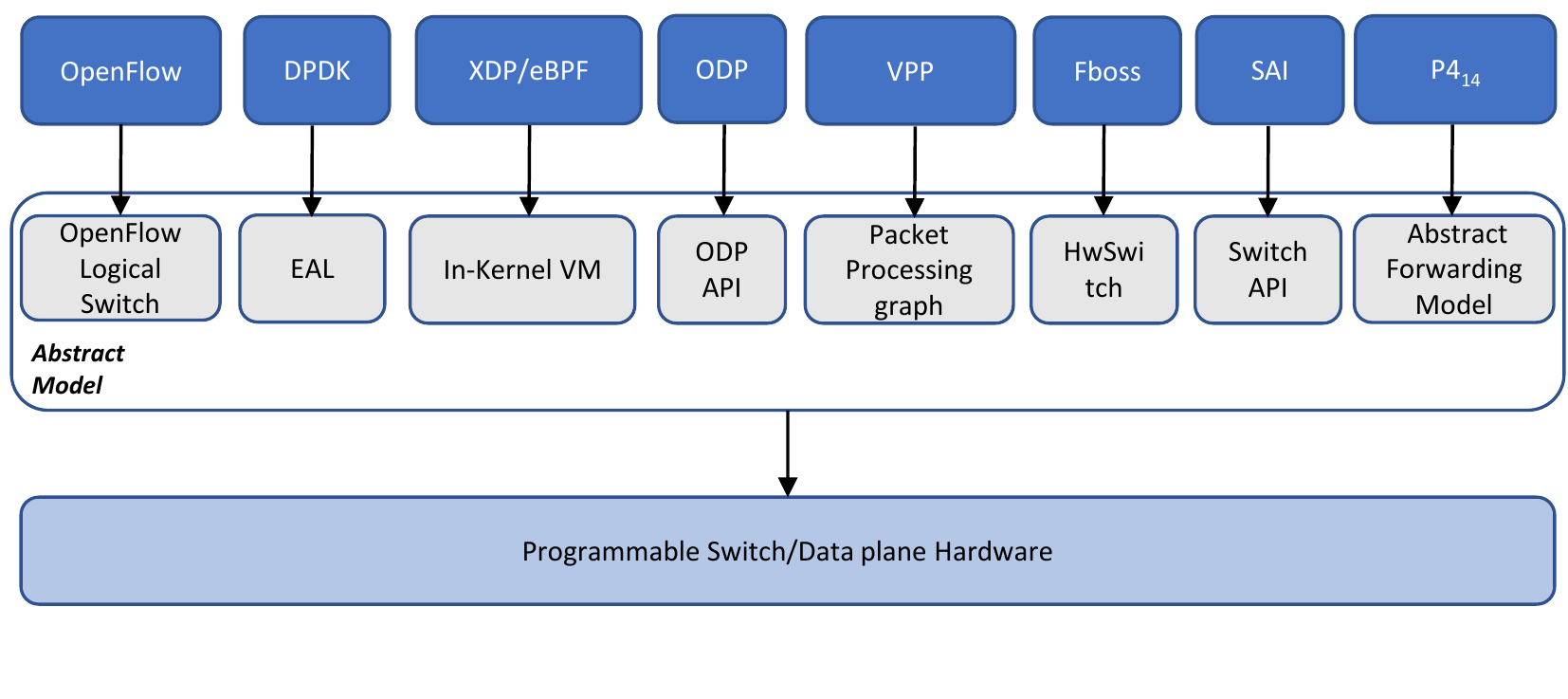}
	\caption{Abstract models used by major SDN protocol (OpenFlow \cite{openFlowSwitchSpecV135}) and programmable data plane technologies (DPDK \cite{dpdk}, XDP/eBPF \cite{hoiland2018express}, ODP \cite{ODP}, VPP \cite{VPP},  Fboss \cite{choi2018fboss}, SAI \cite{switchsai}, P4\textsubscript{14} \cite{p4v14})}
	\label{fig:dpandabslayer}
\end{figure}

Programmable switch needs a software stack for programming the data plane. Abstraction layer  (\textbf{\textit{Device and resource Abstraction Layer  (DAL)}} \cite{rfc7426}) is one of the key component of these stacks. It provides a uniform view of the hardware and a convenient way to develop complex constructs rather than  directly  dealing with actual hardware instructions. Prominent programmable data plane technologies have heterogeneous hardware internals and corresponding hardware abstraction layer  (\textbf{HAL}). They are different from each other (Fig. \ref{fig:dpandabslayer}). But, without a common abstraction layer both data and control plane application become tightly bound to target architecture. For example, a program developed for RMT based architectures can not be executed directly over a DPDK based smart-NIC. Lack of common abstraction layer increases cost and complexity of development, testing, performance bench-marking and formal verification of any novel network functionality built over it. Besides this, a well designed hardware  abstraction layer  (Logical Forwarding Plane \cite{casado2010virtualizing})  is center piece  for successful use of network  virtualization, network function virtualization and service chain composition  \cite{Casado:2014:ASN:2661061.2661063}. Though a common abstraction layer over heterogeneous hardware architecture has various advantages, but restricting to a single abstract model of hardware closes the door for future  innovation in both hardware and abstraction layer design. PDP programming stacks should be decoupled from  hardware architecture and capable of accommodating multiple hardware architecture. 

Majority of the data plane programming stacks are tightly coupled with their own abstraction. Whereas, P4  (dominant data plane programming language) decoupled hardware architecture definition from packet processing behavior in it's latest version   (P4\textsubscript{16} \cite{p4v16}).  P4\textsubscript{16} provides separate language constructs to express abstract hardware architecture and develop program based on those architectures. Taking advantage of this feature, most of the data plane programming stacks have already developed interface with P4 \cite{choi2017pvpp,voros2018t4p4s,tu2018linux,patra2017macsad,switchsai,shahbaz2016pisces}. Besides this, several other architectures are also being proposed \cite{P4targets,psaSpec,Chole:2017:DDP:3098822.3098823,Rim:2018:HPP:3302425.3302435,he2018p4nfv}. PSA \cite{psaSpec}  is one of the most matured among them. But majority of them lacks programmability of two important components: buffer and scheduler. Moreover, these hardware architecture definitions and how they process a packet are described in an informal language. It creates several issues. Most important among them are- \textbf{\textit{I1}}) identifying clear definition of various components of a hardware architecture become hard \textbf{\textit{I2}}) boundary between 2 components and how they connect with each other become unclear \textbf{\textit{I3}}) exact details of how a packet is processed inside a component may differ from one hardware vendor to another \textbf{\textit{I4}}) developing a formal structure or framework for data plane programs become hard. As consequence of \textbf{\textit{I1}}) and \textbf{\textit{I2}}), modular hardware/simulator/test-bed design become hard. As consequence of \textbf{\textit{I3}}) application layer suffers. For example, if 2 hardware architecture not agree about  packet processing states inside the components, taking snapshot of a network or testing/validating a data plane program for multiple architecture become extremely hard. As a result of \textbf{\textit{I4}}), understanding and comparing various PDP devices and their programmability features become difficult. Moreover without a structural notion, formal analysis of data plane programs become hard. To tackle these challenges, a generalized but flexible yet abstract model of the programmable data plane is necessary. It should be modular in nature with well defined structure of the components. Besides this, well defined interface among the components are necessary for independent development and designing new packet processing architectures based on these components. Moreover, there should be a  high level work flow of the components for ensuring uniform behavior across different hardware architectures.

Considering importance of an abstract model and current technology landscape, we think  it is important to start discussion on a better model of abstraction layer  for programmable switches. In this work, we have attempted to shed light on this topic. In doing so, we make a number of important contributions. After discussing background and related state of the art (in sec. \ref{RelatedWork}), we provided a unified definition of programmable switch (in sec. \ref{ProgrammableSwitch}) which is independent of any protocol stack. Then we presented the design of \textbf{\textit{AVS}}, an example model of \textbf{\textit{Device and resource Abstraction Layer (DAL)/Hardware Abstraction Layer (HAL)}} for programmable data plane devices (in sec. \ref{AbstractModel}). It is modular in design and components are defined with a uniform and structured  functional interface. In discussing details of the components (in sec. \ref{AVSComponents}), at first a generic structure for the components with well defined type of pogrammability features (\textbf{\textit{Compile Time Programmability (CTP) and Run Time Configurability (RTP)}}) are laid out. Then a detailed analysis of structure, workflow and progammability features of each components are discussed (in sec. \ref{PacketHeaderDefinitionSubSection} - \ref{EgressPort}). To present their workflow in a unambiguous and hardware independent manner, we have followed EFSM approach. After discussing components of \textbf{\textit{AVS}} in details, a novel approach to compare programmability level of selected set of PDP devices based on programmability featues of \textbf{\textit{AVS}} components are also presented (in sec. \ref{ProgrammabilityFeatureMatrix}). We have also presented 4 important use cases of \textbf{\textit{AVS}} (in sec. \ref{MotivatingScenarios}).

We do not claim the novelty of underlying components design. We are influenced by several existing work on this domain ~\cite{Bosshart:2013:FMF:2486001.2486011,6665172,Mittal:2015:UPS:2834050.2834085,8117533,Lin:2018:PRN:3185467.3185473}. We believe our achievement is providing a broader picture with an improved and structured abstract model for the programmable switches.

\section{Background and Related Works} \label{RelatedWork} 
OpenFlow \cite{McKeown:2008:OEI:1355734.1355746} protocol is the most successful name in SDN paradigm. It is described over an abstract model of switch named \textbf{\textit{`OpenFlow Logical Switch'}} \cite{openFlowSwitchSpecV135}. OpenFlow decoupled control plane  (CP) and data plane  (DP).  It provides limited programmability in data plane by controlling DP behavior from CP through dynamically configuring a broad set of protocol header fields. Soon academia and industry realized that, with it's ever growing protocol field set OpenFlow can not achieve the goal of true programmable network. True potential of SDN can be leveraged only if data plane is fully programmable.

After the initial wave of OpenFlow  based  switches, work on programmable data plane device has gained significant momentum \cite{bifulco2018survey,kaljic2019survey}. The first consolidated effort for data-plane programmability  can be attributed to RMT \cite{Bosshart:2013:FMF:2486001.2486011}, which proposed an architecture for programmable data plane devices. Based on RMT's architecture, an \textit{`abstract forwarding model'} for data plane devices has been presented in \cite{Bosshart:2014:PPP:2656877.2656890}. In this work, authors have also presented a data plane programming language  named P4 \cite{Bosshart:2014:PPP:2656877.2656890} to write program for RMT based programmable switches.  Based on success of RMT and P4, several commercial programmable data plane devices has been emerged in the market \cite{BarefootTofino,agilioCX}. In initial version  (version 14) of P4   (\textbf{P4\textsubscript{14}}) \cite{p4v14}, data plane programs (DPP) were developed based on  \textit{`abstract forwarding model'}  of  ~\cite{Bosshart:2014:PPP:2656877.2656890}.  P4\textsubscript{14} was strictly coupled with the proposed abstract forwarding model. Although P4\textsubscript{14} came out with an abstract model and related programming construct to program them, it's high cohesion with proposed hardware with limited programmability made it unsuitable for design of new data plane hardware architectures. This clearly shows the requirement of  dis-aggregation between hardware architecture and programming language for programmable data plane devices.

Currently  RMT is the dominant programmable data plane architecture.  But it has few limitations. each pipeline stage of RMT architecture can only access memory allocated for it. As a result cross component access of memory/data is also not possible \cite{Chole:2017:DDP:3098822.3098823}. Moreover, as components of RMT pipeline are connected  linearly, a packet can not skip unnecessary stages in the pipeline. To overcome these limitations, authors have proposed dRMT architecture in \cite{Chole:2017:DDP:3098822.3098823}. It improves RMT architecture by dis-aggregating memory and compute resources inside switch.  In  \cite{Rim:2018:HPP:3302425.3302435} authors presented a conceptual model of data plane for supporting parallel processing. It is heavily based on RMT and dRMT architecture. But these works, do not provide concrete hardware abstraction layer.

Efforts for making server based networking environment more programmable have also seen massive growth. DPDK \cite{dpdk} and eBPF/XDP \cite{hoiland2018express} are the two major framework for userspace and in-kernel packet processing. They rely on two abstraction layers: EAL for DPDK and in-kernel virutal machine for eBPF/XDP. Though they are able to work in conjunction with smart-NIC but they are not suitable for core switches. Another important category of programming stack is API based abstraction layer \cite{ODP,SAI}. Instead of providing an abstraction layer they provide an API through which data plane can be programmed. These APIs not provide any guideline about how to implement a feature. All conforming hardware vendors provide their own implementation of the API. As a result \textbf{\textit{I3}}  (sec. \ref{Introduction}) is not solved by API based solutions. Moreover to interface with data plane programming language  (ex. P4), it needs more than one step. At first step, P4 codes are transformed to abstract API calls and then API calls are transformed to hardware specific instructions. Thus this style abstraction increases complexity.

Among all those data plane programming technologies, P4 and relevant tool-sets emerged as the most dominant programming stack. Nearly all the major programmable data plane  platform have a P4 implementation or P4 interfacing \cite{choi2017pvpp,voros2018t4p4s,tu2018linux,patra2017macsad,switchsai,shahbaz2016pisces}. As P4 is becoming more matured, several works \cite{188956,Dang:2017:WPL:3050220.3050231,Hancock:2016:HUP:2999572.2999607,8038396,Liu:2018:PPV:3230543.3230582,Notzli:2018:PAT:3185467.3185497,Abhashkumar:2017:PPO:3050220.3050235,patra2017macsad,he2018p4nfv} focusing on various high level aspects of data plane is rolling out. But, as all of these works  depends on P4, they are limited by the abstract forwarding model of P4\textsubscript{14}  or few of the present hardware model \cite{P4targets} supported by P4\textsubscript{16}.

Initial version of P4  (${P4}_{14}$) was strictly coupled with the proposed \textit{`abstract forwarding model'} and RMT architecture. But later, P4 community have taken a clean state approach and followed  the proven path of virtual machine world. In current version   (version 16) of P4  (\textbf{P4\textsubscript{16}}) \cite{p4v16} language constructs for expressing data plane device architecture and hardware specific features are decoupled from the constructs for expressing packet processing logic.  Currently in a P4\textsubscript{16} based data plane program, developer needs to define a model of hardware  (in form of include file) and the actual packet processing logic separately. This decoupling gives 4 crucial advantage: a) new hardware architecture  can be supported any time b) common hardware architecture or abstraction layer can be created  by all relevant stake holders c) top down design of programmable data plane hardware become possible d) high level language constructs for packet processing can be developed independently from hardware design . As a result, P4 can keep the path open for hardware innovation yet promoting interoperability through common hardware architectures. On the other hand P4\textsubscript{16} enables writing data plane behavior in a flexible and portable manner based on abstraction layers.

P4 community is actively working toward developing a common hardware architecture that can cover various programmable data plane devices ranging from core switches to smart-NIC. PSA \cite{psaSpec} is the  most matured attempt from P4 community toward that goal. It aims to list several common packet processing paths inside programmable switch and smart-NIC. These paths are composed from   multiple programmable P4 blocks. It also lists several related stateful data structure and  functions for use in data plane. PSA can be identified as the first  structured effort toward a standard  abstraction layer of data plane. But PSA specification describes the components and their work flow in a descriptive way. As a result there is always chance of confusion among various implementation of PSA and it suffers from problem \textbf{\textit{I1,I2,I3}}  (section \ref{Introduction}). Besides, PSA has no option of programmablity for  buffer ~\cite{Lin:2018:PRN:3185467.3185473}  and scheduler ~\cite{Mittal:2015:UPS:2834050.2834085,Sivaraman:2015:TPP:2834050.2834106} in the architecture.  In this work, we intend to  overcome these limitations by proposing a better abstract model  (\textbf{\textit{AVS}}) for programmable data plane devices by extending PSA architecture.

Our approach is aligned with P4 community, as the  proposed abstract model  (\textbf{\textit{AVS}}) is inspired by PSA and P4.  It also enhances the \textbf{\textit{PSA}} architecture, firstly by adding programmability to  buffer and scheduler  components. Secondly, by providing a common functional structure for each components and expressing work flow of each components as hardware  architecture agnostic extended finite state machine. These enhancements can help in designing various innovative applications (sec. 8 \cite{AVSTechReport})). Moreover, P4 doesn't impose any restriction on use of hardware architecture and it has decoupled hardware dependent compiler back end for various architectures. Hence, use of \textbf{\textit{AVS}} like abstract model doesn't restrict use of other hardware architectures and the scope of innovation in hardware design remains wide open.

\section{Programmable Switch} \label{ProgrammableSwitch}
\begin{figure*}
\includegraphics[ width=7in,trim={0.0in .6in 0in 0.0in},clip]{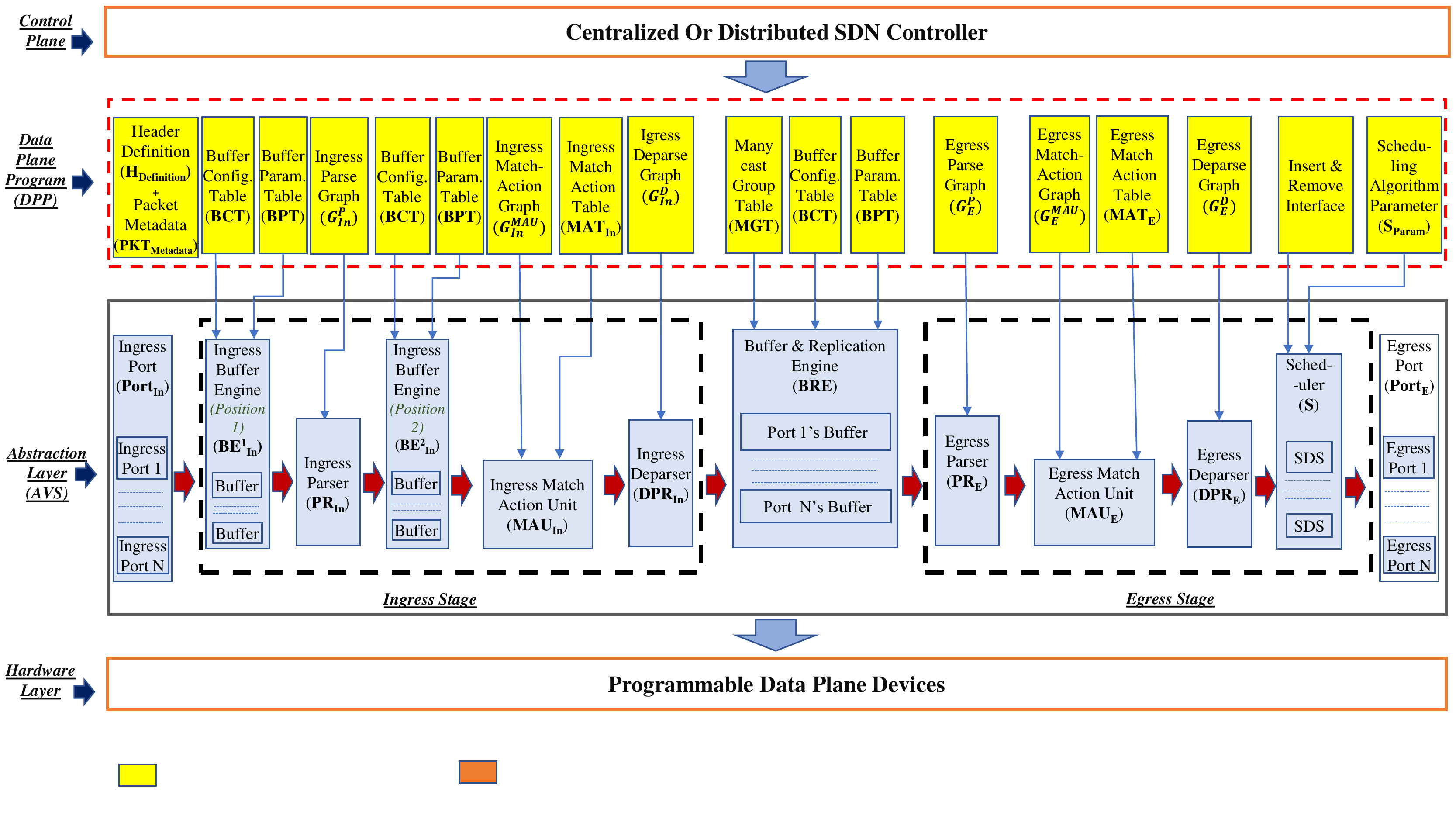}
\caption{An Abstract Model for Programmable Switch}
\label{fig:abstractModel}
\end{figure*}

A packet  ($PKT$)'s  life  inside a switch starts with reaching through incoming  (ingress) port as few bits of data and ends with exiting through outgoing  (egress) port. 
$$PKT = \{{Bit}_{1},{Bit}_{2},{Bit}_{3},............,{Bit}_{Packet \: Length}\}$$
How a data plane device acts  is defined by 
\begin{itemize}
  \item \textbf{\textit{S1- }Interpreting Packet:} how to interpret incoming set of bits  ($PKT$) as different meaningful fields. 
  \item \textbf{\textit{S2- }Packet Metadata:} a data plane device keeps few hardware dependent information about a packet  (metadata) for  use in different stages of the packet's life cycle. Example: arrival time of a packet, this is necessary for packet scheduling. 
  \item \textbf{\textit{S3- }Packet Processing Work-flow:} Set of operations executed based on different fields of packet and various data structures.
  \item \textbf{\textit{S4- }Control Plane Configuration Parameters:}  data plane  lacks of global knowledge about the network. Control plane needs to configure parameters so that \textbf{\textit{S3}} can be adapted with  dynamic  network conditions.
  \item \textbf{\textit{S5- }Emitting Packet:} What's the structure of the packet emitted by the device.  
\end{itemize}

In legacy switches, \textbf{\textit{S1-S3} \& S5} are static.  Though \textbf{\textit{S4}} is available, they are rather configuring parameters for fixed \textbf{\textit{S3}}. Hence, in legacy switches, once \textbf{\textit{S1-S5}} are loaded, they can't be changed. These switches are optimized for specific protocol. But in programmable switch,  \textbf{\textit{S1-S5}} are not static and they can be modified at switch lifetime.  Formally, any combination of hardware and/or software is a \textbf{\textit{programmable switch or programmable data plane device}}, if it fulfills following properties
\begin{itemize}
\item \textbf{\textit{P1:}} Not bound to any specific protocol 
\item \textbf{\textit{P2:}} Can execute any program, which
	\begin{itemize}
		\item \textbf{\textit{P2-1:}} contains logic and interface for \textbf{\textit{S1-S5}} 
		\item \textbf{\textit{P2-2:}} is not coupled to hardware architecture
		\item \textbf{\textit{P2-3:}} can be loaded and unloaded at runtime
	\end{itemize}
\end{itemize}
Any logical abstraction that can provide an uniform view and functionality of a \textit{`programmable switch'} can be termed as an \textbf{\textit{abstract model}} of programmable switch.

\section{The Abstract Model  (\textbf{\textit{AVS}})} \label{AbstractModel}

\begin{figure}
\centering
\includegraphics[ width=\columnwidth, trim={0.0in 0.8in 7in 0.0in},clip]{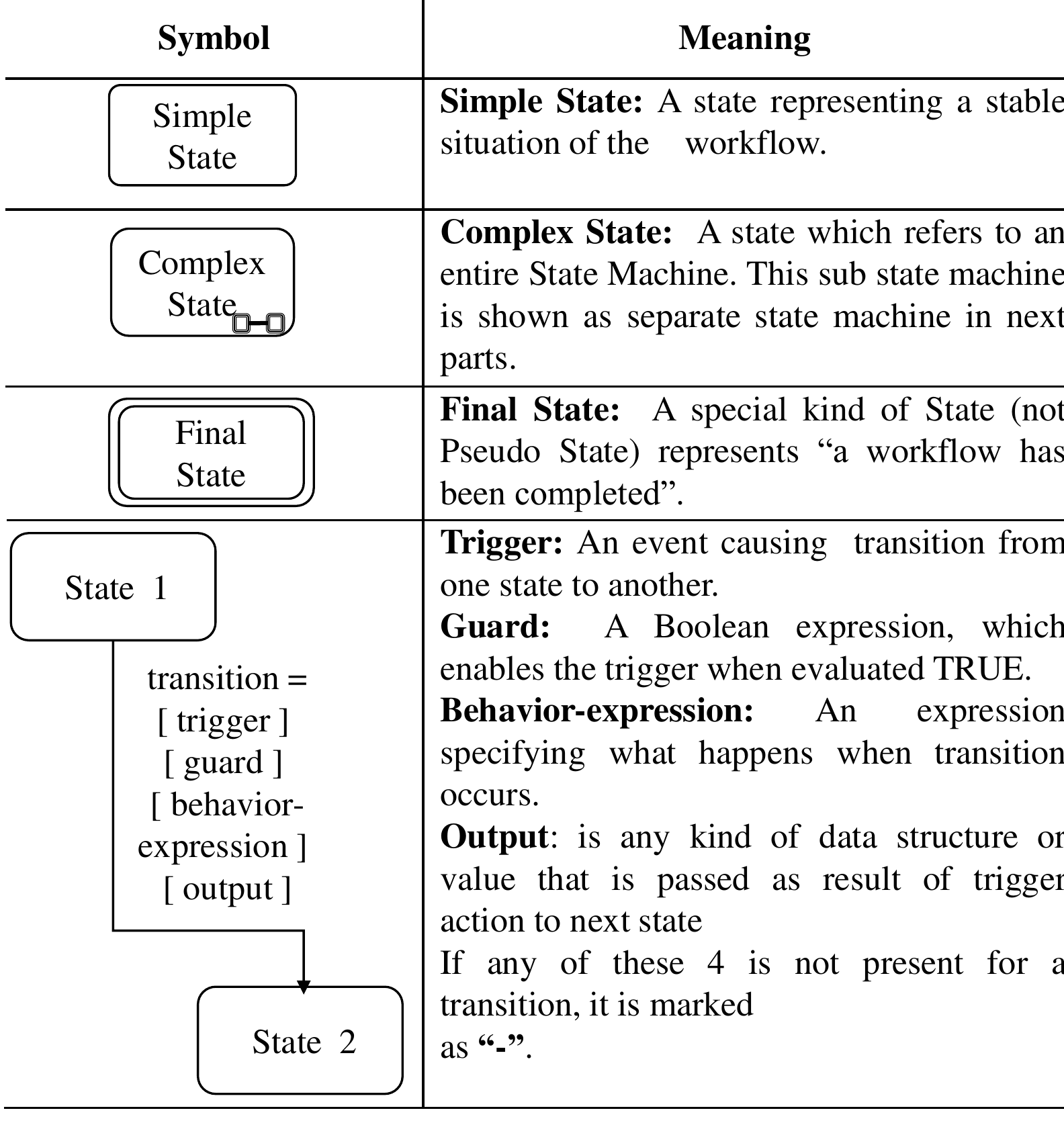}
\caption{EFSM Notations}
\label{fig:fsmNotations}
\end{figure}

Processing path of a packet inside a programmable switch  can be modeled as an abstract pipeline of serially  connected programmable components.  Our proposed pipeline based abstract model of programmable switch  (\textbf{\textit{`Abstract Virtual Switch  (AVS)'}}) is represented along with other components of SDN stack  in  Fig. ~\ref{fig:abstractModel}. Components  ($\mathcal{C}$) of \textbf{\textit{AVS}} are a) ingress port  (${Port}_{In}$) b) ingress parser  (${PR}_{In}$) c) ingress buffer engine  (${BE}_{In}$) d) ingress match action unit  (${MAU}_{In}$) e) ingress deparser  (${DPR}_{In}$) f) buffer and replication engine  (${BRE}$) g) egress parser  (${PR}_{E}$) h) egress match action unit  (${MAU}_{E}$) i) egress deparser  (${DPR}_{E}$)  j)  scheduler  ($S$) k) egress port  (${Port}_{E}$).



Based on egress port selection for a packet, the \textbf{\textit{AVS}} pipeline is logically divided into 2 stages: \textbf{ingress stage}- before selecting the egress port and \textbf{egress stage}- after selecting egress port. After entering egress stage a packet's egress port can not be changed. This is particularly important when packets are replicated  (clone, broadcast, multicast etc.) in egress stage. On such cases, packets may require further match action processing in egress stage. For example, traffic rate controlling at each outgoing port requires action at egress stage. If egress processing is not required in a hardware implementation, vendors may skip that part.



\textbf{\textit{AVS}} may represent a single non virtualized programmable switch or a slice in a virtualized programmable switch or just a software switch.  It is compiler's duty to map components of \textbf{\textit{AVS}} to actual hardware resources. \textbf{\textit{AVS}} provides a uniform view of data plane   over heterogeneous programmable switches. How the data plane will behave is defined by  \textbf{data plane program  (DPP)}  (Fig. ~\ref{fig:abstractModel}). Management plane handles  (un)loading of DPP. On the other hand control plane controls runtime behavior of \textbf{\textit{AVS}} by configuring parameters.

\section{AVS Components  } \label{AVSComponents}
\subsection {Generic Structure Of The Components} \label{genericStructureOfComponents}
Each component  ($\mathcal{C}$) of \textbf{\textit{AVS}} represents a programmable unit and they are needed to be programmed from outside before a packet processing starts. These are supplied as DPP.  On the other hand CP controls each component's behavior by configuring parameters through southbound interface. Degree of programmability of \textbf{\textit{AVS}} and it's components depend on following 2 kind of features.
\begin{itemize}
	\item \textbf{\textit{Compile Time Programmability  (\textbf{CTP}) Features}}: Set of instructions a programmable component can execute  (comparable to cpu instruction set). How $\mathcal{C}$ will behave at run time is defined through these features. 
	\item \textbf{\textit{Run Time Configurability  (\textbf{RTC}) Features}}: Capability of adjusting run-time behavior of \textbf{CTP} features through configuring parameters.  Control plane uses these to manage the behavior of a component $\mathcal{C}$. 
\end{itemize}

Irrespective of hardware implementation, CTP and RTC features can be exposed to upper layer as an uniform API. Compiler  translates these API call to actual hardware instruction. DPP is a program expressed through CTP features and contains runtime processing logic of a component $\mathcal{C}$. And control plane application controls behavior of those processing logic at run time through $RTC$ features. DPP also contains data structures for  facilitating control plane communication with $\mathcal{C}$ through RTC features. 

Formally, a component  ($\mathcal{C}$) can be represented as a component function $f_c$,  
\begin{equation} \label{functionalDescriptionOfCompo}
{f}_{c}: \mathcal{I} \rightarrow \mathcal{O}  \\
{f}_{c}  (X,ProcLogic,{Conf}_{param}) = Y
\end{equation} 

Here,
\begin{itemize}
	\item $\mathcal{I}$ is the domain of ${f}_{c}$, it represents the set of all possible  values that  $\mathcal{C}$ accepts.
	\item $\mathcal{O}$ is the co-domain  (range) of ${f}_{c}$, it represents the set of all possible  values that $\mathcal{C}$ can return as output. 
	\item $ProcLogic$ is the processing logic to be executed by the component. It is represented using $CTP$, $RTC$ and control flow
	\item $X$ is the input  to the component, $X \in I$
	\item $Y$ is the output of  the component, $Y \in O$. If $\mathcal{C}$ modifies $X$ and returns the result as variable of $X$, $Y = X*$ notation is used. Ex. ingress match action unit (sec. \ref{IngressMatchActionUnit})take $PHV$ as input and returns result as a modified $PHV$.
	\item ${Conf}_{param}$ is the set of parameters \textbf{\textit{CP}} can configure to control the behavior of the component at run time. This is a subset of $CTP$ features 
	\end{itemize}

\begin{figure*}
\centering
\includegraphics[ width=\textwidth,trim={0.0in 0in 0in 0.0in},clip]{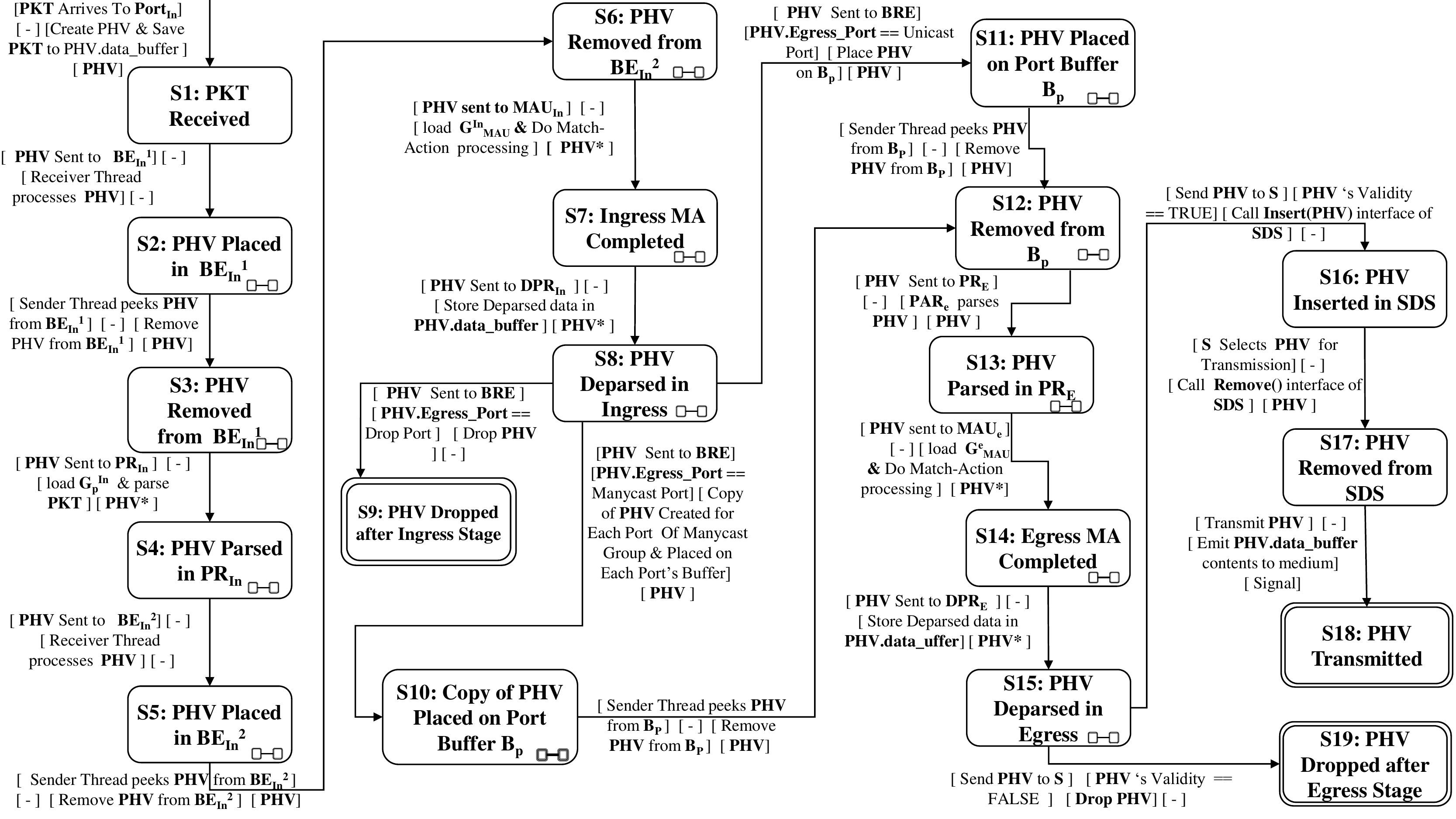}
\caption{Packet's Life Cycle EFSM}
\label{fig:main_state_machine}
\end{figure*}
In next few subsections, each component of \textbf{\textit{AVS}} are discussed in details. How they work and their programmability features are also discussed. For expressing work-flow of the components extended finite state machine  (EFSM) approach is used. In  figure ~\ref{fig:main_state_machine}, an EFSM is presented  to define how a packet  (\textbf{\textit{PKT}}) goes through different components of the \textbf{AVS}. Separate sub state machines are used when necessary. Table ~\ref{substateMachineList} contains complex states of this state machine (figure ~\ref{fig:main_state_machine}) and reference to corresponding sub-state-machine with section number where the components are discussed. Notations used in the EFSMs are described in  Fig. ~\ref{fig:fsmNotations}. Parser, match action unit and deparser  exists both in \textit{`ingress'} and \textit{`egress'} stage. Their internal structure and parameters structures are same for both the stage but parameter names are different. To differentiate between ingress and egress stage components and parameters   \textbf{\textit{In}} and \textbf{\textit{E}} subscripts are used  respectively.

\begin{table}
\caption{Complex state of Fig. ~\ref{fig:main_state_machine} and corresponding sub-state-machines}
\label{substateMachineList}
\begin{tabular}{lll }
\toprule
State & Sub statemachine  & Section \\
\midrule
S2,S5, S10,S11 & Fig. ~\ref{fig:bufferRcvr} & Section ~\ref{bufferSubSection} \\
S3,S6,S12 & Fig.  ~\ref{fig:bufferSender}&  Section  ~\ref{bufferSubSection} \\
S4,S13 & Fig.   ~\ref{fig:parserStateMachine}&  Section  ~\ref{IngressParserSubSection}, ~\ref{EgressParser}\\
S7,S14 & Fig.  ~\ref{fig:mat}&  Section ~\ref{IngressMatchActionUnit}, ~\ref{EgressMatchActionUnit} \\
S8 ,S15 & Fig.  ~\ref{fig:deparser} &  Section ~\ref{IngressDeparserSubSection}, ~\ref{EgressDeparser}\\
\bottomrule
\end{tabular}
\end{table}

\subsection{Related Terminology}  \label{PacketHeaderDefinitionSubSection}
Before driving into  details of each component, few relevant  terminologies and notations are discussed in this section.

\subsubsection{Bit Space  ( $\mathcal{BS}$)}
Upon receipt a packet  ($PKT$) is a sequence of one and zero.
$$PKT = \{{Bit}_{1},{Bit}_{2},{Bit}_{3},............,{Bit}_{Packet\, Length}\}$$
Formally, a packet  ($PKT$) is a point in the space \\ ${BS} = {\{0,1\}}^{Maximum\, Packet\, Length}$.  

\begin{figure}
\centering
\includegraphics[width=\columnwidth , trim={0.0in 5.05in 6.5in 0.0in},clip]{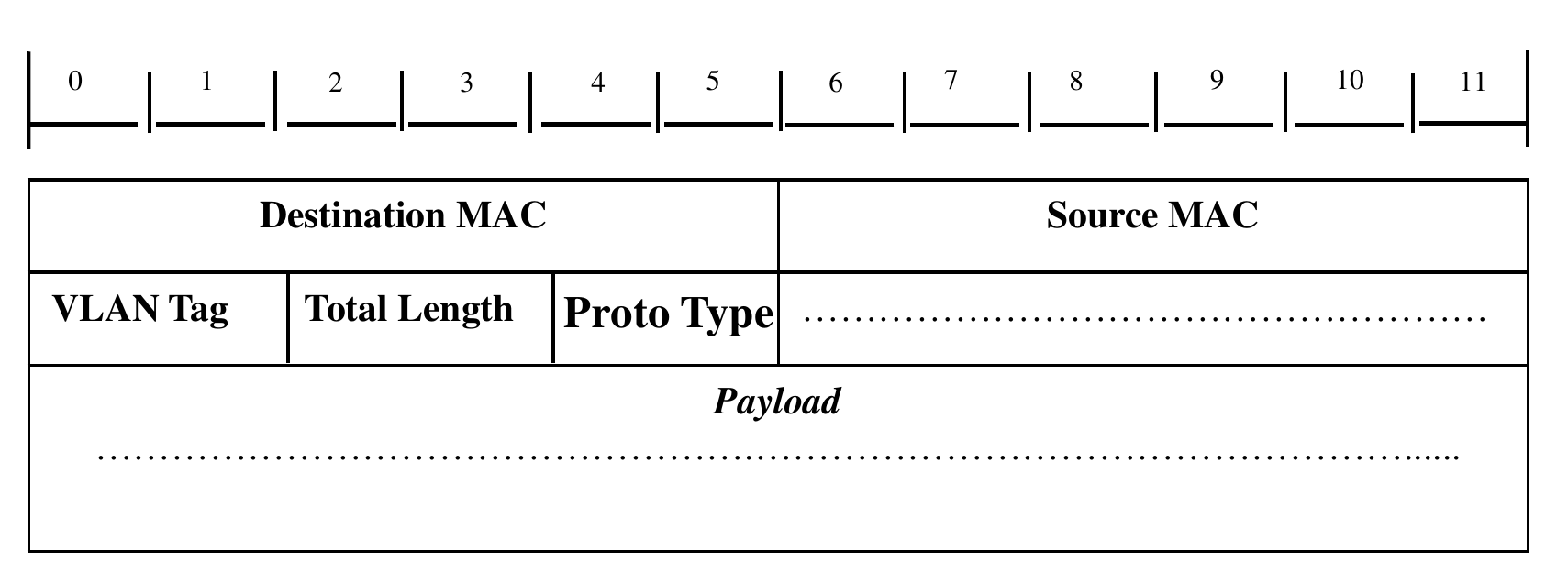}
\caption{A sample packet format}
\label{fig:headerSample}
\end{figure}

\subsubsection{Header Definition  (${H}_{Definition}$)} \label{HeaderDEfinition}
To do meaningful operation on a $PKT$, it is needed to be interpreted as fields of different protocols. Header definition provides structure of these fields. Definition of i'th header field 
$${H}_{f}^{i} =  (unique\, ID/name,\, starting  \, position \, in \, packet,\, length) $$
Here,
$$ \: \:   (1\leq i \leq p) \, \, and $$
$$p \, = \: total \: number \:  of \: fields \; in \: {H}_{Definition} $$
 Starting position of  ${H}_{f}^{i}$ provides  relative order of the field in packet. All  ${H}_{f}^{i}$ together from packet header,
$${PKT}_{header} = \bigcup_{i=1}^{p} H_{f}^{i}$$
Rest of the packet is considered as payload. 

\subsubsection{Packet Metadata  (${PKT_{Metadata}}$)} \label{PacketMetadata}
Depending on actual hardware implementation, switches maintain some metadata   (${PKT}_{Metadata}$) about a packet  (ingress or egress port, time of arrival, packet unicast or multicast type etc.). I'th metadata field 
$${M}_{f}^{i} =  (unique ID/name,\, data \; type/length)$$
Here,
$$ \: \:  (1\leq i \leq q) \, \, and $$
$$q \, = \: total \: number \:  of \: fields \; in \: {PKT_{Metadata}} $$

\begin{figure}
\centering
\includegraphics[width=\columnwidth , trim={0.0in 1.5in 6.5in 0.0in},clip]{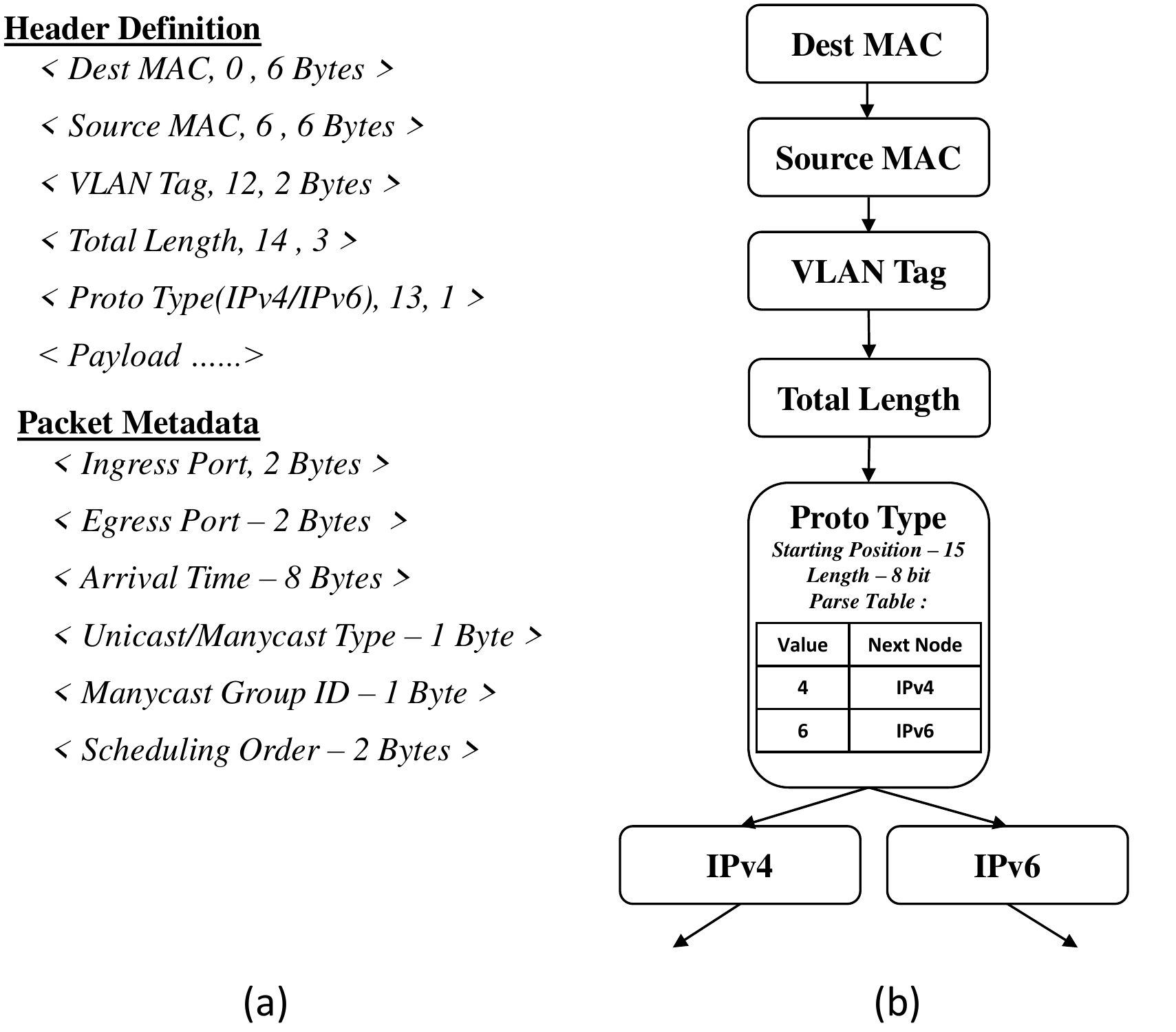}
\caption{ (a) Header definition and metadata for the sample packet of Fig. ~\ref{fig:headerSample}  (b) Parse graph for sample packet of Fig. ~\ref{fig:headerSample} (only "Proto Type" node is shown in expanded format) }\label{fig:HeaderDefinitionAndPArseGraph}
\end{figure}

All  ${M}_{f}^{i}$ together from packet metadata,
$${PKT}_{Metadata} = \bigcup_{i=1}^{q} {M}_{f}^{i}$$
A sample packet format is shown in Fig.~\ref{fig:headerSample}, corresponding header definition \& a packet metadata are shown in Fig.~\ref{fig:HeaderDefinitionAndPArseGraph}. (a).

\subsubsection{Packet Header Vector  (PHV) and Space  (PHVS)} \label{PAcketHeaderVEctor}
Header definition and metadata defines a $p+q$ dimensional space named Packet Header Vector Space  (PHVS). Each point in this space is termed as packet header vector  (PHV). PHV can be considered as a container for all the attributes of a packet in key-value format. Where, key represents a header field  ($H_f$) or  metadata field   (${M}_{f}$) or any other field derived in the pipeline and value represents corresponding data. 
$$PHV = {PKT}_{header}  \: \cup  {PKT}_{Metadata}$$
Each PHV represents a point and flow represents a region in  PHVS. In a PHV, all the fields in definition may not be present at some point in AVS. They may be filled up by different components in pipeline at different stage of the packet's life-cycle. 

\subsubsection{Ordered PHV Set  (${PHV}_{set}, <$)} \label{OrderedPHVSet}
Let ${PHV}_{set}$ is a set of $PHV$. An \textit{`Ordered PHV Set'} is an ordered pair   (${PHV}_{set}, \leq $) of set ${PHV}_{set}$ and the binary relation $\leq$ contained in ${PHV}_{set} \times {PHV}_{set}$, such that
\begin{itemize}
\item \textbf{Reflexive: } $\forall PHV \in {PHV}_{set} : PHV \leq PHV$
\item \textbf{Transitive: }$\forall {PHV}_{i}, {PHV}_{j}, {PHV}_{k} \in {PHV}_{set} : [  ({PHV}_{i} \leq {PHV}_{j}) \:  \&\& \:   ({PHV}_{j} \leq {PHV}_{k})] \:$ \\ $ \Rightarrow   (  ({PHV}_{i} \leq {PHV}_{k})) $
\item \textbf{Anti-symmetry: }$\forall {PHV}_{i}, {PHV}_{j}\in {PHV}_{set} :$ \\$ [  ({PHV}_{i} \leq {PHV}_{j}) \: \&\&   ({PHV}_{j} \leq {PHV}_{i})] \: \Rightarrow$ \\ $   (  ({PHV}_{i} = {PHV}_{k})) $
\end{itemize}
Here the relation is defined based on one or more fields of $PHV$ such that, for a field 
$x \in PHV$ and 2 elements of the set ${PHV}_{m}, {PHV}_{n} \in {PHV}_{set}$,
$  ({PHV}_{m}.x.value \, \leq \, {PHV}_{n}.x.value) \Rightarrow   ({PHV}_{m} \, \leq \, {PHV}_{n})$
. If $({PHV}_{m}.x.value \,$ \\$ == \, {PHV}_{n}.x.value)$ then another filed $y \in PHV$ can be used to break tie. As packets received from each port have distinct arrival time, and each port has different number in a switch, a total order on a set of $PHV$ is always possible.

\textbf{\textit{Header Definition}} \& \textbf{Packet Metadata  (${PKT_{Metadata}}$)} together forms PHVS.  Metdata is fixed for an architecture. PHVS is mainly dependent on Header definition. Nearly all the component's of \textbf{\textit{AVS}} domain and range is PHVS. Hence it is a crucial part of DPP.

\section{Analysis of Components} \label{ComponentsSection}

\subsection{Ingress Port  (${Port}_{In}$)}  \label{IngressPortSubsection}

In \textbf{\textit{AVS}}, sole task of an \textbf{\textit{ingress port  (${Port}_{In}$)}} is to receive a set of bits  ($PKT$) and store as a PHV. How a hardware level frame is received and format of the frame is out of the scope of our discussion. After receiving, a new $PHV$ is initiated and $PKT$ is stored in  \textbf{\textit{data\_buffer}} variable of PHV. \textbf{\textit{data\_buffer}} is a storage for an array of bits. In this phase, necessary metadata  (ingress port, arrival time etc.) are also stored in the PHV.  Then the $PHV$ is  passed to next component in pipeline. In AVS, ingress ports do not have any kind of programmability in terms of $CTP$ and $RTC$ features. 
\begin{gather*}
f_{{Port}_{In}}: {BS} \rightarrow {PHVS}  \\
f_{{Port}_{In}}  (PKT,Null,Null) = PHV , \, 
\end{gather*}

\subsection{Ingress Parser  (\textbf{\textit{${PR}_{In}$}})} \label{IngressParserSubSection}
\begin{figure}
\centering
\includegraphics[ width=\columnwidth , trim={0.0in 0in 6.5in 0.0in},clip]{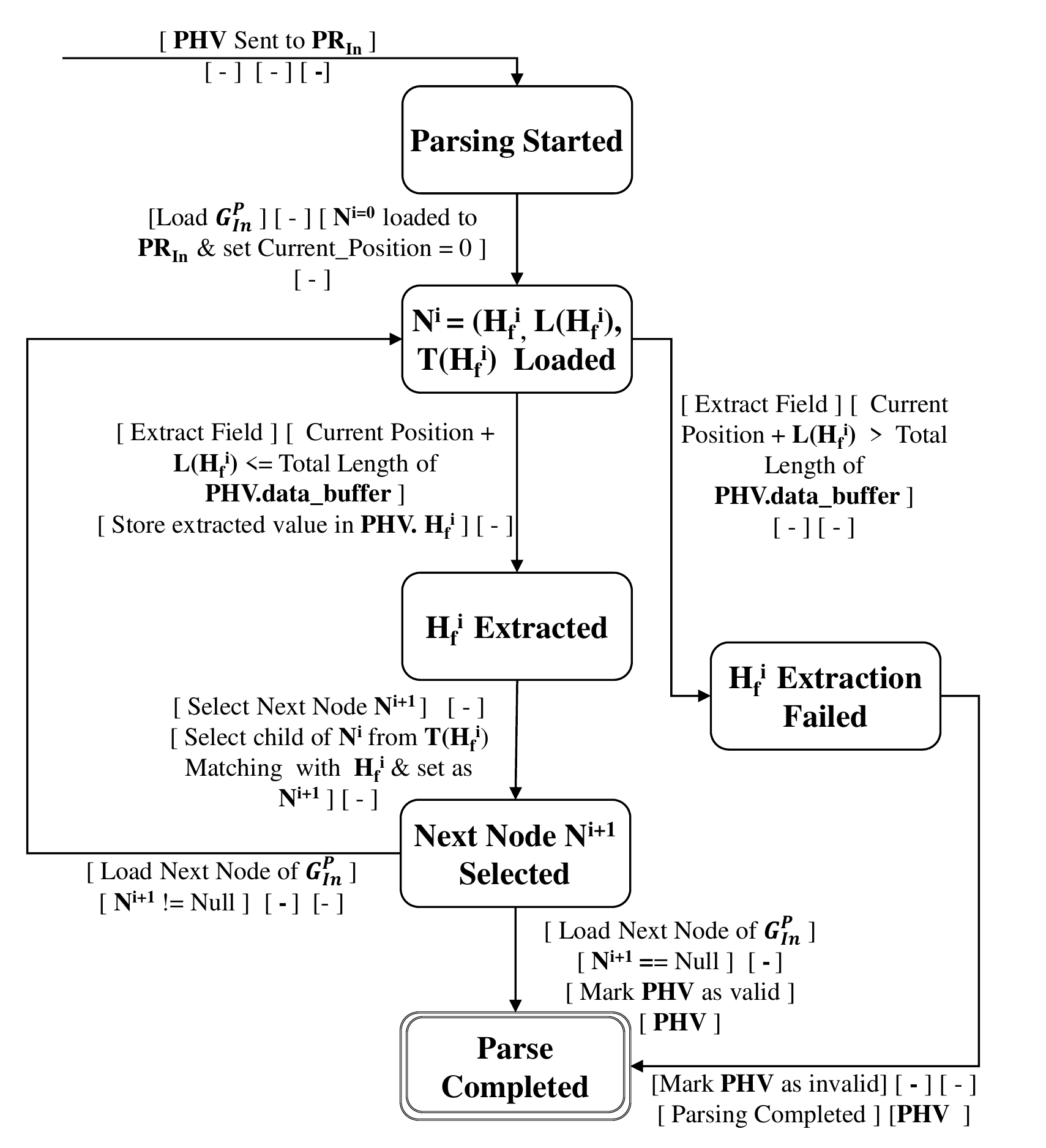}
\caption{Programmable Parser EFSM}
\label{fig:parserStateMachine}
\end{figure}

${PR}_{In}$ parses array of bits stored in $PHV.data\_buffer$ to different header fields. Parsing logic is provided through parse graph  ($G^{p}_{In}$). Each node  ($N^{i}$) $\in$ $G^{p}_{In}$ contains: a) Header field information: ${H}_{f}^{i} \in PHV$ to be parsed, starting position in $PKT$  and length of ($L  ({H}_{f}^{i})$)  and b) Parse table ~\cite{6665172} (  ($T  ({H}_{f}^{i})$): Lists possible values of ${H}_{f}^{i}$ and corresponding next node. Work-flow of a programmable parser is shown in Fig.~\ref{fig:parserStateMachine} as EFSM. After parsing, PHV is passed to next component in pipeline. 
\begin{gather*}
f_{{PR}_{In}}: {BS} \rightarrow {PHVS}  \\
f_{{PR}_{In}}  (PHV.data\_buffer,G^{p}_{In},Null) = PHV^* 
\end{gather*}

\textbf{CTP features}: a) maximum length limit: maximum length of $PKT$ that can be parsed by ${PR}_{In}$. This limit is important for bounded runtime.    b) supported data types: different data types  (including all the standard  primitive data types: int, float, char etc) that can be supported by the parser c) granularity of field parsing: can the parser circuit parse individual bit or variable number number of bits to a field.

\textbf{RTC features}: a) modifiability of $G^{p}_{In}$ :  is the parse graph  (${G_{p}}^{In}$) and it's nodes contents are modifiable at runtime by CP.

\subsection{Ingress Buffer Engine  (\textbf{\textit{${BE}_{In}$}})}   \label{bufferSubSection}

General role of a buffer is to temporarily hold packets. Though a programmable buffer can be expressed under match-action semantics, but its significance in various important applications \cite{Lin:2018:PRN:3185467.3185473,kogan2017programmable} warrants a separate component for buffer. Ingress buffer engine  (${BE}_{In}$) consists of a set of buffer  (${B}_{set}$). Assuming $n$ individual buffers  ($B_{i}$) in the engine, ${B}_{set}\, = \,  \bigcup_{i=1}^{n} B_{i} $. Their size may be fixed or programmable. 
CP can control these sizes through configuring \textbf{\textit{`Buffer Parameter  Table   ($BPT$)'}}  (Table ~\ref{sampleBPTTable}). There are 2 possible positions of ${BE}_{In}$ in  AVS. Any one or both  can be used. 

First one  ({${BE}_{In}^{1}$}) is just after the ${Port}_{In}$. \textbf{${BE}_{In}^{1}$} holds PHV  received from ports. CP controls  PHV from which ingress port should be stored to which buffer through \textbf{\textit{`Buffer Configuration Table  (\textbf{${BCT}$})'}}.

Second one  ({${BE}_{In}^{2}$}) is after ingress parser. This is a more generalized and useful implementation which can  store PHV to a buffer based on PHV fields. Here, instead of only ingress port, CP configures based on which header/metadata field, a PHV should be sent to which buffer through \textbf{\textit{`Buffer Configuration Table  (\textbf{${BCT}$})'}}  (Table ~\ref{sampleBCTTable}). Priority for matching PHV's also can be assigned through ${BCT}$.

\begin{figure}
\centering
\includegraphics[ width=\columnwidth , trim={0.0in 0in 6.5in 0.0in},clip]{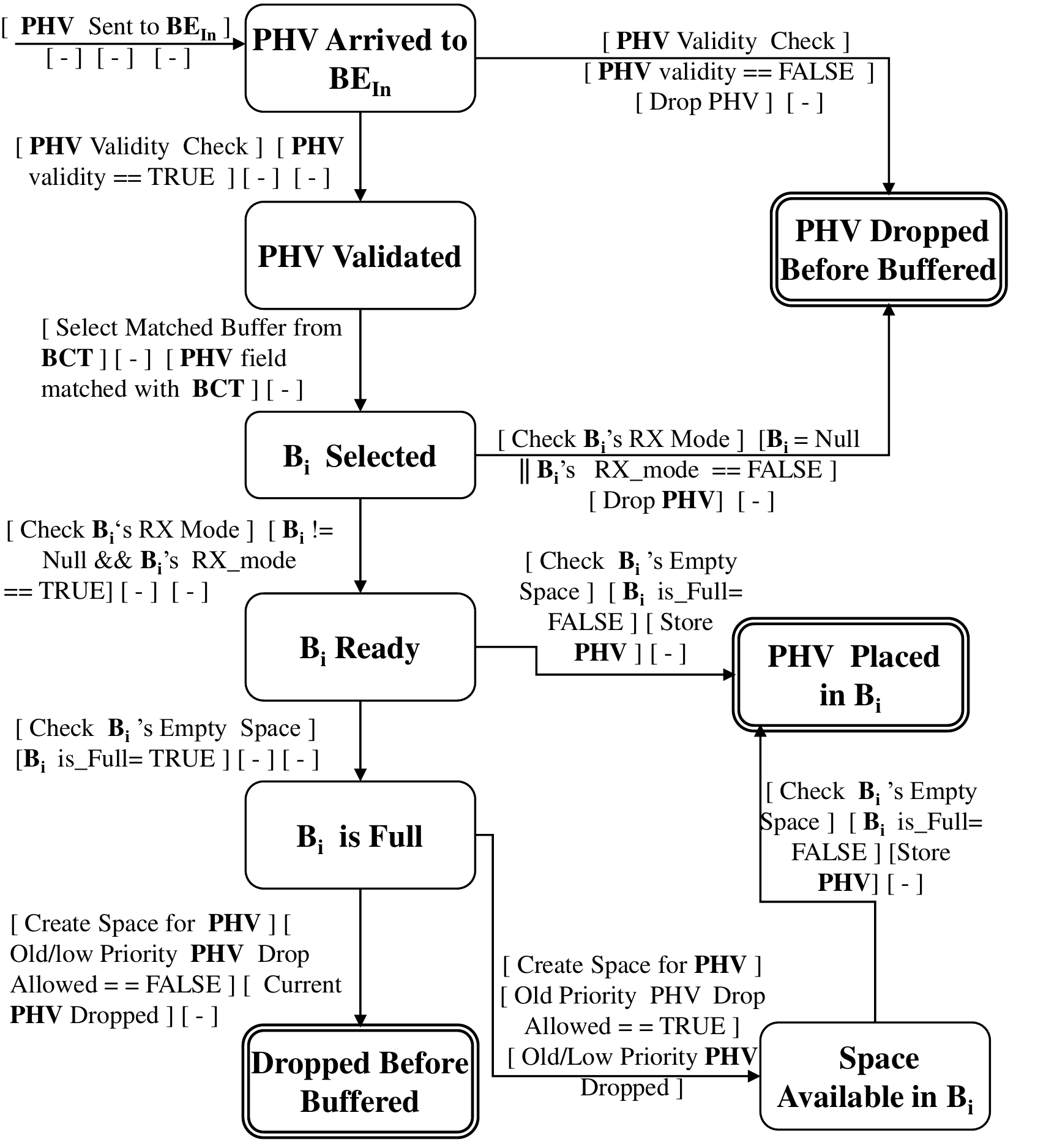}
\caption{Buffer Receiver Thread EFSM }
\label{fig:bufferRcvr}
\end{figure}

\begin{figure}
\centering
\includegraphics[ width=\columnwidth , trim={0.in 1.6in 6.5in 0.0in},clip]{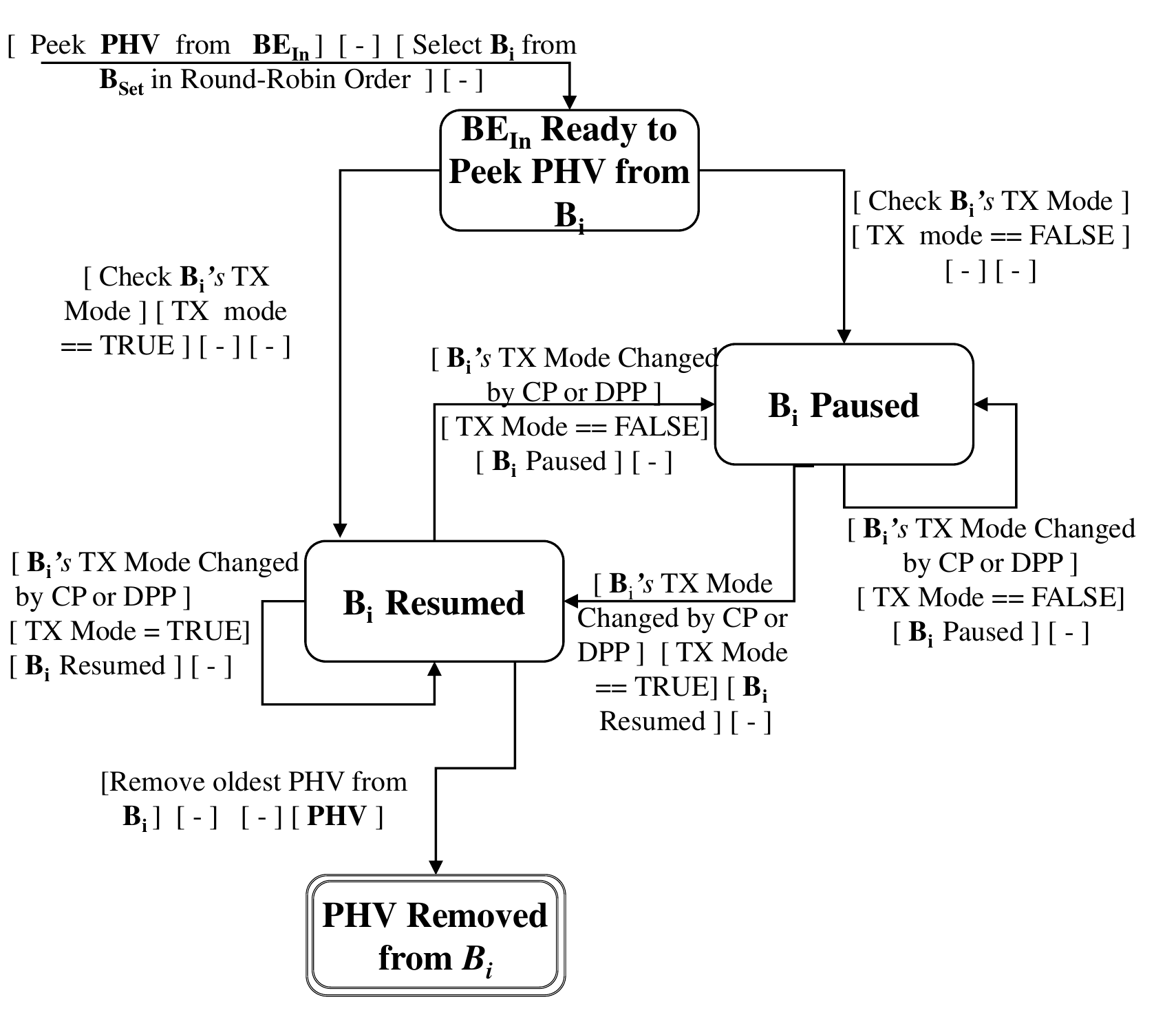}
\caption{Buffer Sender Thread EFSM }
\label{fig:bufferSender}
\end{figure}

Buffer engine's work-flow can be expressed as 2 threads: a) \textbf{\textit{Receiver Thread}} (Fig.~\ref{fig:bufferRcvr}): for inserting PHV in buffers and b) \textbf{\textit{Sender Thread}} (Fig.~\ref{fig:bufferSender} ): for moving out PHV from buffers and sending to next component in pipeline. These 2 threads behavior  are controlled by 2 separate state variable for each of the buffer  (${B}_{i}\in B_{Set}$). These are configurable from control plane via \textbf{\textit{`Buffer Parameter Table  (\textbf{${BPT}$})'}}. These 2 are: a) RX mode:  when RX of a ${B}_{i}$ is true, receiver thread either receive PHV and store them (from port or ingress parser) or drop them if false b) TX mode:  hold flow from buffer  (buffer in pause state) when TX is false or release  (buffer in resume state) packet/PHV to next component in pipeline when TX of the buffer  (${B}_{i}$) is true.

\begin{table}
\caption{Buffer Configuration Table (BCT) for sample packet of Fig. ~\ref{fig:headerSample}}\label{sampleBCTTable}
\begin{tabular}{llll}
\toprule
PHV  field Name & PHV  field Value &  Buffer ID & Priority \\
\midrule
VLAN Tag & 0x 00 15 25 &  3  & 1 \\
VLAN Tag & 00 45 25   & 6   & 0\\
$\cdots$ & $\cdots$ & $\cdots$ & $\cdots$\\
\bottomrule
\end{tabular}
\end{table}

\begin{table}
\caption{Buffer Parameter Table (BPT) for sample packet of Fig. ~\ref{fig:headerSample}}
\label{sampleBPTTable}
\begin{tabular}{llll }
\toprule
Buffer ID & Size &  RX & TX\\
\midrule
\midrule  
1 & 2048 &  True & False  \\
5 & 3072   & False & True   \\
$\cdots$ & $\cdots$ & $\cdots$ & $\cdots$\\
\bottomrule
\end{tabular}
\end{table}

Receiver thread of buffer engine can be expressed as 
\begin{gather*}
f_{{BE}_{receiver}}: PHVS \rightarrow {B}_{set} \\
f_{{BE}_{receiver}} (PHV, Null, \{BPT , BCT\}) = {B}_{set}; \,  (PHV \\
\: \: \: \: \: \: \: \: \: \: \: \: \: \: \: \: \: \: \: \:\: \: \: \: \:\: \: \: \: \: inserted \: in \: {B}_{set})
\end{gather*}

and sender thread as (assuming a round-robin order for selecting next buffer from where next PHV will be peeked)

\begin{gather*}
f_{{BE}_{sender}}: {B}_{set} \rightarrow PHVS \\
f_{{BE}_{sender}} (Null, Null, Null) \: =PHV 
\end{gather*}

\textbf{CTP features}: a) buffer size \& number controlling: number of buffers and corresponding size are programmable or not  b) $BCT$ creation: is $BCT$ fixed or can be declared (size and definition) in compile time c) global access of buffer property: can other components of $AVS$ access buffer properties?

\textbf{RTC features}: a) BPT modifiability: can $CP$ manipulate buffer size through $BPT$ b) $BCT$ modifiability: can $CP$ add, modify or delete entry in $BCT$ table.

\subsection{Ingress Match Action Unit (\textbf{\textit{${MAU}_{In}$}})} \label{IngressMatchActionUnit}
This component can be considered as the computational unit of a data plane device. What is cpu in a server, match-action unit can be considered as that unit for programmable data plane devices. Increasingly, more complex and generalized processing units (FPGA, CPU, GPU) are being proposed as the main computational unit for PDP devices. but match-action base hardware are still dominating.  In this work, we have expressed the generic packet header based computations in match-action semantics. hardware implementations may use CPU/GPU/RMT circuit for implementing the actual logic. But the concept is generalized for the component. 

In this component, values of $PHV$  field or any other data derived from them  are matched with either a) control plane configured data or b) data collected by data plane itself.  Based on matching result different actions are executed. Control plane configured data are kept in a table like data structure. Control plane can store, modify or delete data from these tables at runtime through southbound API. Data plane's collected(or derived) data can be of 3 types a) stateful information about a flow(meter, register, counter etc.) b) stateless metadata about a packet and c) any constant value supplied at compile time in DPP. 

Programmability and performance of a switch mostly depends on the set of actions it allows programmer to use. Actions can be of different types : \textbf{stateless} - only access and modify current PHV fields, \textbf{stateful} - access and update previously stored data about a flow and use them to to update PHV. 

Different data plane programming language may represent match action semantics in different syntax. But fundamentally, match-action block requires 4 information, a) name/id of the PHV field which will be matched/compared b) control or data plane supplied value, against which PHV field values are matched or compared c) matching/comparison method (exact, ternary, <, > , != etc.) d) one or more action(action block) to  execute based on comparison result. 

\begin{table}\footnotesize{\small}
	\caption{A sample MAT for \textit{`Proto Type'} field in sample packet of Figure ~\ref{fig:headerSample}}
	\label{tab:sampleMAT}
	\begin{tabular}{ccc}
		\toprule
		Match Type & Values to be Matched &  Action[s]\\
		\toprule
		\midrule  
		Exact & 4    &  Increase IPv4 counter   \\
		Exact & 6    &  Drop Packet  \\
		$\cdots$ & $\cdots$   & $\cdots$  \\
		\midrule  
		\bottomrule
	\end{tabular}
\end{table}

Without loosing generality, here we assumed that, storing computational logic for packet processing can be represented as graph. In current literature, generic data structure for storing processing logic information is termed as $Match-Action-Table(MAT)$ \cite{Bosshart:2013:FMF:2486001.2486011}. Processing logic for a match-action-unit can be represented as a match-action graph (${G}^{MAU}_{In}$), where each node($N^i $) represents match-action logic for a specific field in PHV and  edges are control flow. Let, a protocol field $p\_f \in PHV$, it's value in PHV is $p\_f.value$ and  data structure for storing  it's match-action information is ${MAT}_{p\_f}$. 
${{MAT}_{p\_f}}  =  \{match \, type \, \times \,values \, to \, be$\\ $ matched \, with \, p\_f \times \,  action[s]\}$. Corresponding $N^i $ is a tuple $(p\_f, {{MAT}_{p\_f}} )$. Result of match-action processing for each field can be stored in stateful memory in case of  stateful operations. Or they can simply update some fields in PHV(ex. updating destination port   according to routing table). An example $MAT$ for \textit{`Proto Type'} field of sample packet (Figure ~\ref{fig:headerSample}) is presented in Table ~\ref{tab:sampleMAT}.

Let, ${{MAT\_SET}}_{In}$ = Set of  $MAT$ for all the nodes($N^i $) in ${G}^{MAU}_{In}$. Formally ,
\begin{gather*}
f_{{MAU}_{In}}: {PHVS} \rightarrow {PHVS}  \\
f_{{MAU}_{In}}(PHV,{G}^{MAU}_{In},{{MAT\_SET}}_{In}) = PHV^*
\end{gather*}

\begin{figure}
\centering
\includegraphics[ width=\columnwidth , trim={0.0in 0in 6.5in 0.0in},clip]{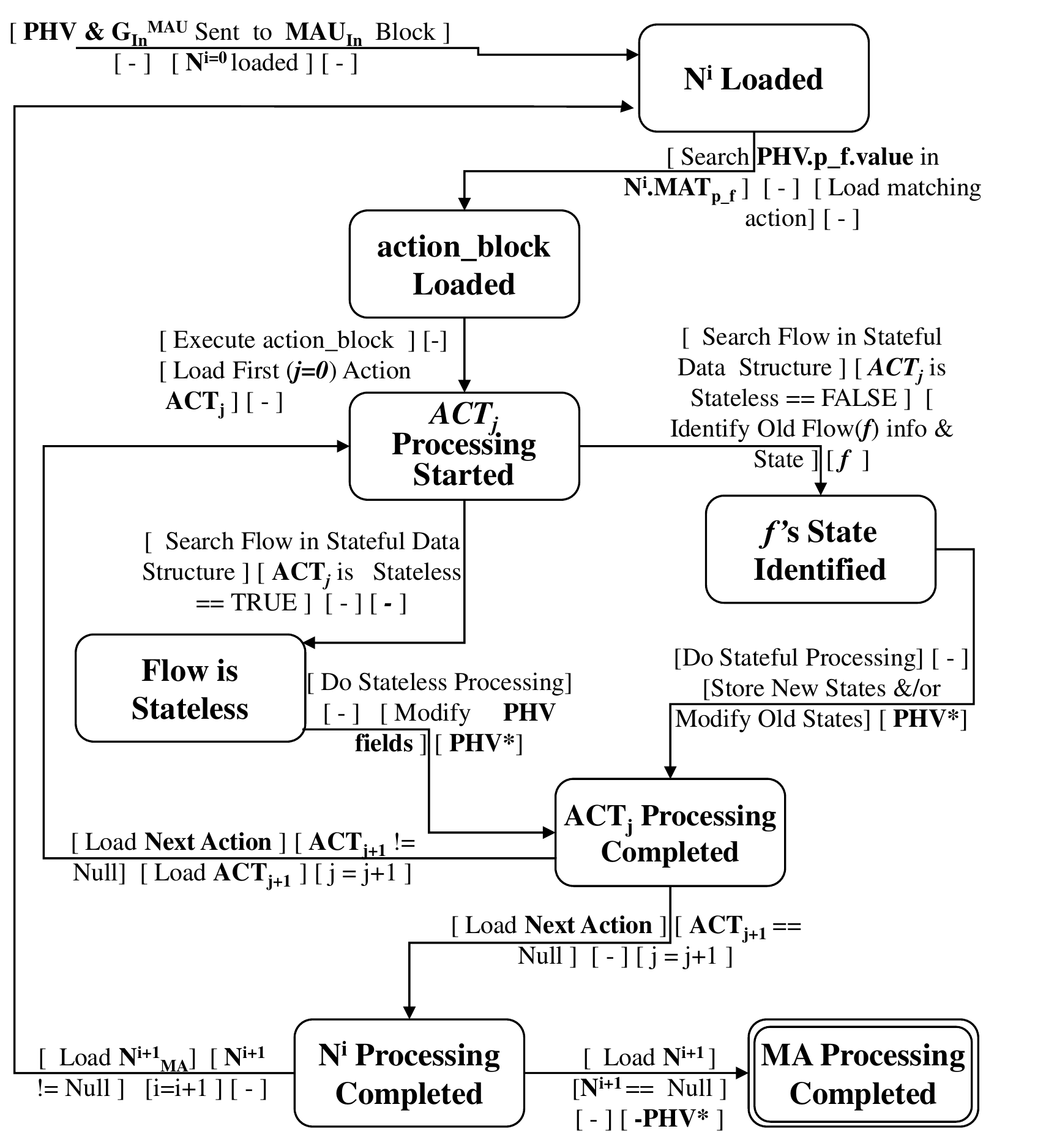}
\caption{Ingress Match Action EFSM}
\label{fig:mat}
\end{figure}

Work-flow of ingress match action unit(\textbf{\textit{${MAU}_{In}$}}) is presented in Fig. ~\ref{fig:mat}.

\textbf{CTP features}: 
a) data type support in MAT:  data types (int, float, bit pattern, string etc.) allowed for lookup in MAT b) matching type support in MAT: type of look ups  (longest prefix match, ternary, exact, range etc.) are supported c) MAT size \& number controlling: MAT size are fixed length or their size can be declared in compile time d) availability of  stateful action: is it possible to maintain state information about the flows e) availability of  stateful data structure: what are the data structures for keeping state information about the flows  (register, counter etc.). Custom data structure can be created or not f) state sharing among different components: are the flow states shareable among different components of AVS. g) MAT modifiability from DPP: is the MAT tables modifiable from the data plane program. This is necessary for advanced and faster decision making in data plane \citep{Turkovic:2018:FNC:3229574.3229581}. h) available action types: what are the different type of actions that can be done on the fields of a PHV.

\textbf{RTC features}: 
a) MAT controlling: can the CP add/ remove/update match-action tables entries b) stateful data access from CP: can the CP read/write stateful data of a flow from/to the DP.

\subsection{Ingress Deparser(\textbf{\textit{${DPR}_{In}$}})} \label{IngressDeparserSubSection}

\begin{figure}
\centering
\includegraphics[ width=\columnwidth , trim={0.0in 1in 6.5in 0.0in},clip]{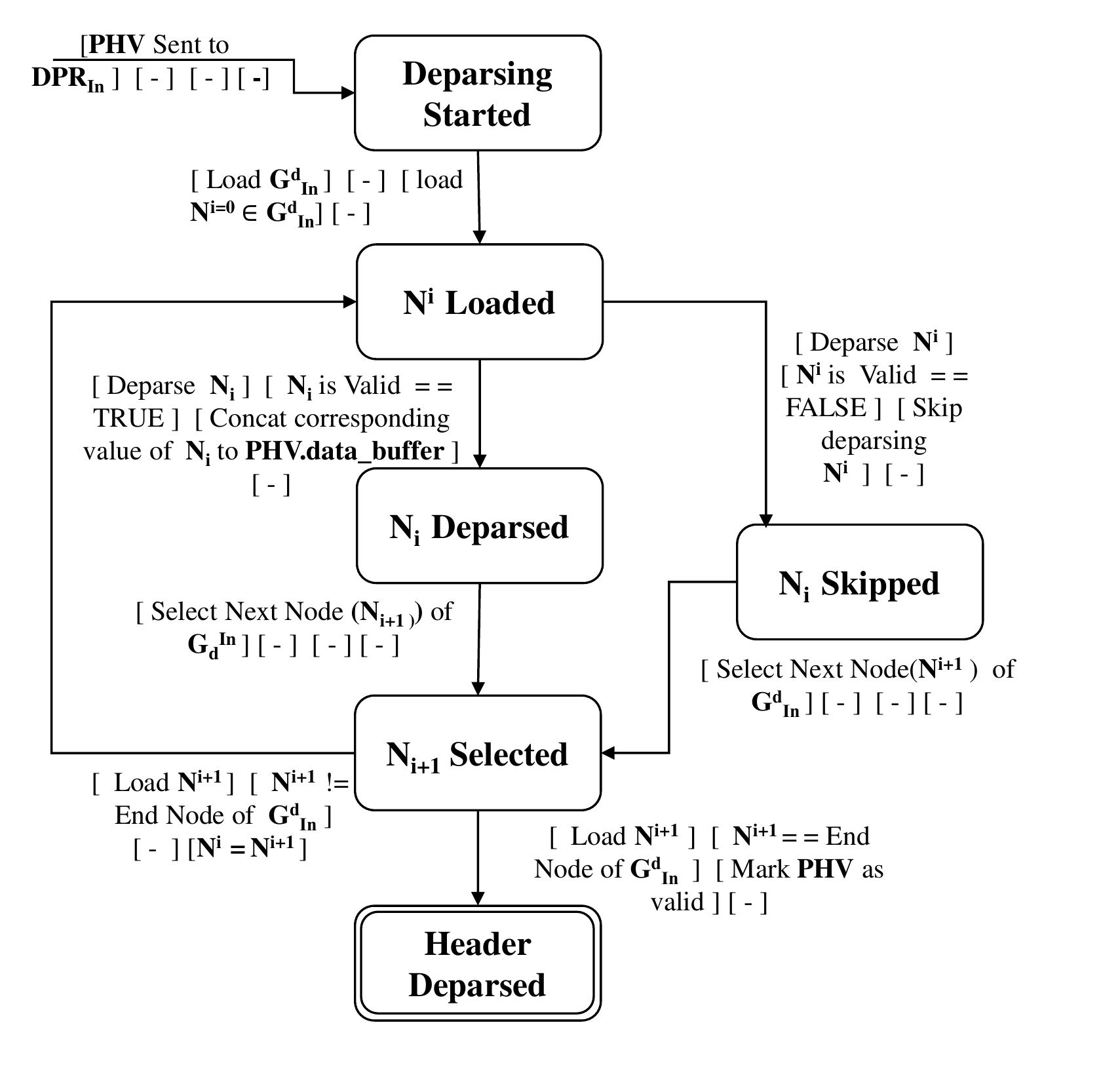}
\caption{Ingress Deparser EFSM}
\label{fig:deparser}
\end{figure}

This component deserializes  a $PHV$. All the fields of a $PHV$ may not be necessary for emitting to next component. Some may be dropped or some may be included more than once. Moreover all the header fields declared in $PHV$ may not be valid for all the packets. In deparser definition, it is defined which of the valid header fields will be included in the outgoing packet. Besides the $PHV$ fields, constant data also can be included in the packet. Deparser definition is transformed to a graph(${{G_d}}^{In}$) in compilation stage. In this graph each node represents a field of $PHV$ or any arbitrary data to be emitted. ${DPR}_{in}$ checks validity of each node and concatenates the field data in \textbf{data\_buffer}(a per packet buffer for storing bits). This also inherently defines each  fields relative position in the packet to be emitted. After deparsing all the nodes of ${{G_d}}^{In}$, payload is concatenated to the \textbf{data\_buffer}. Work-flow of an ingress deparser unit is represented as EFSM in Figure ~\ref{fig:deparser}. 
\begin{gather*}
f_{{DPR}_{in}}: {PHVS} \rightarrow {PHVS}  \\
f_{{DPR}_{in}}(PHV, {{G}}^{In}_d, Null)  = PHV^* 
\end{gather*}

\textbf{CTP features}: 
a) maximum length limit: maximum length of a packet that can be created through the deparser unit b) data type support: data types that can be emitted  by the deparser. 

\textbf{RTC features}: a) ${{G}}^{In}_d$ modifiability: can CP modify ${{G}}^{In}_d$ at run time.


\subsection{Buffer and Replication Engine($BRE$)} \label{BufferAndReplicationEngine}
$BRE$ is the bridge between ingress and egress stage. It contains buffer  for each egress port. These buffers can be just simple FIFO buffer or fully programmable buffer (sec. \ref{bufferSubSection} ). But header/metadata based buffer assignment of programmable buffer (${BE}_{In}^{2}$ section ~\ref{bufferSubSection}) is not needed here, because we assume BRE buffers will be reserved as per port resource. Moreover, adding extra programmable circuit increases the packet processing delay. 
$$BRE.{B}_{set}\, = \,  \bigcup_{i=1}^{n} B_{i} \:, n \,= total \, number \, of \, egress \, port$$

For a $PHV$, egress port selection is done in ingress match action unit. After ingress deparser, the PHV may be destined toward either a single egress port(\textbf{unicast}) or  more than one egress  port (\textbf{manycast - multicast, broadcast}).  
\begin{itemize}
	\item For unicast port, PHV is simply moved to egress port's buffer with all of its metadata.
	\item For manycast port, control plane needs to configure group membership of the ports through \textit{`Manycast Group Table(MGT)'}. BRE will make necessary number of copies of the  PHV for each member of the group.   Those PHV are  placed to buffer of relevant egress ports. 
\end{itemize}

Insertion of $PHV$ to port buffer ($B_p$) are handled by receiver thread of buffer. Sender thread of buffer removes $PHV$  from port buffer($B_p$) and either drops or passes to egress stage. Work-flow of these 2 threads are same as of ingress buffer engine discussed in section ~\ref{bufferSubSection}.

Formally 
$$f_{BRE}: {PHVS} \rightarrow {{PHVS}^{n}} $$
$$f_{BRE}(PHV, Null, MGT) = n \, copies \, of \, PHV$$
Here, n = total  number  of  egress  port  in a manycast  group. For unicast packets n = 1.


\textbf{CTP features}:  a) all  of the $CTP$ features of ${BE}_{in}$ (except $BCT$ )

\textbf{RTC features}: a) all  of the $RTC$ features of ${BE}_{in}$ (except $BCT$  configuration) b) Manycast group membership table configuration: can CP modify manycast group and their members at runtime.

\subsection{Egress Parser (${PR}_{E}$)} \label{EgressParser}
After PHV are removed from $BRE$, they are sent to egress parser. Structurally it is same as ingress parser.  Only difference is, after finishing egress parsing of packet header, instead of sending to a buffer, PHVs are sent to egress match-action unit. 

$$f_{{PR}_{E}}: {BS} \rightarrow {PHVS}  $$
$$f_{{PR}_{E}} (PHV.data\_buffer,G^{p}_{E},Null) = PHV^* $$

\subsection{Egress Match Action Unit (${MAU}_{E}$)} \label{EgressMatchActionUnit}
Same as ingress match action. Only restriction is, egress port can not be changed in this stage. 
Formally ,
\begin{gather*}
f_{{MAU}_{E}}: {PHVS} \rightarrow {PHVS}  \\
f_{{MAU}_{E}} (PHV,{G}^{MAU}_{E},{{MAT\_SET}}_E) = PHV^*
\end{gather*}

\subsection{Egress Deparser (${DPR}_{E}$)} \label{EgressDeparser}
Same as ingress deparser. Only difference is, after egress deparsing a PHV is sent to scheduler instead of buffer and replication engine.
\begin{gather*}
f_{{DPR}_{E}}: {PHVS} \rightarrow {PHVS}  \\
f_{{DPR}_{E}} (PHV, G_E^d, Null)  = PHV^* 
\end{gather*}

\subsection{Scheduler($\mathcal{S}$)} \label{Scheduler}
After egress processing is complete, a $PHV$ is passed to Scheduler($\mathcal{S}$) for transmission  to next hop. Scheduling defines \textbf{order} and \textbf{time} \cite{Sivaraman:2015:TPP:2834050.2834106} of a PHV's transmission. To achieve these goals, PHV is needed to be stored in specified order (according to scheduling algorithm) inside an appropriate scheduler data structure (\textbf{\textit{SDS}}) (i.e. Queue, collection of queue, tree etc.). Different hardware implementation can use different data structure. From an abstract point of view, \textbf{\textit{SDS}} maintains an ordered PHV Set (Section ~\ref{OrderedPHVSet}). From \textbf{\textit{SDS}} PHV's are selected according to scheduling algorithm for transmission at appropriate time. As there is no universal scheduling algorithm ~\cite{Sivaraman:2013:NSB:2535771.2535796} to match  all kind of application goals, there is no common implementation of scheduler unit also.

To create a common and generalized abstract model,  2 abstract interfaces are assumed in conjunction with SDS. Different scheduling algorithm will implement these 2 interfaces differently based on data structures provided by the hardware. CP can adjust behavior of $S$ through configuring scheduling algorithm parameters(${S}_{param}$). Example: table for configuring weights of weighted fair queuing algorithm. These 2 abstract interface are following:
\begin{itemize}
	\item \textbf{Insert(PHV)}: Implementation of this interface  calculates the relative order of a $PHV$ inside \textbf{\textit{SDS}}  and places it in corresponding position. Let, $SchedulingOrder$ is the metadata field that contains the PHV's order in $SDS$. Now assume, ${PHV\_SET}^{'}$ is a set of  $PHV$ ordered by $SchedulingOrder$ (in reality this order may depend in more than one field) stored in $\textbf{SDS}$. 
	\begin{gather*}
	{PHV\_SET}^{'} = \{{PHV}_{1},{PHV}_{2},{PHV}_{3},..\\
	\: \: \: \: \: \: \: \: \: \: \: \: \: \: \: \: \: \: \: \: \: \: \: \: \: \: \: \:\: \: \: \: \: \: \: \:  ..........,{PHV}_{n}\}
	\end{gather*}
	Finding a new PHV(${PHV}_{new}$)'s scheduling order requires computation of ${PHV}_{new}.SchedulingOrder$  from following types of data. 
	
	\begin{itemize}
		\item value in ${PHV}_{new}$'s  header fields($H_f$) and/or  metadata (${M}_{f}$) 
		\item stateful information about the flow, computed and stored by switch 
		\item CP configured parameter such as weight for a weighted fair queuing scheduling
	\end{itemize}
	
	Insert interface provides a total order (Section ~\ref{OrderedPHVSet}) on $({PHV\_SET},\: <)$ based on $SchedulingOrder$ 
	where, 
	\begin{gather*}
	PHV\_SET =  {PHV\_SET}^{'}\cup \{{PHV}_{new}\} \\ 
	({PHV\_SET}, <) = \{{PHV}_{1},{PHV}_{2},{PHV}_{p},\\
	\: \: \: \: \: \: \: \: \: \:   .....{PHV}_{new},{PHV}_{q},  .................,{PHV}_{n}\}
	\end{gather*}
	Therefore,  insert interface can be functionally represented as 
	\begin{gather*}
	f_{{S}^{insert}}: {PHVS} \rightarrow {Ordered \: PHV \: Set(SDS, \leq)}\\
	f_{{S}^{insert}}(PHV, Insert\,Interface\,Implementation, \, \\
	S_{param})= {PHV\_SET}
	\end{gather*}
	Computing ${PHV}_{new}.SchedulingOrder$  can be done either in match-action stage and result can be carried to scheduler unit through PHV. Or also can be computed in scheduler unit alone.
	After that, scheduler will insert the packet at appropriate location in the data structure. In case of storage shortage a low priority packet can be dropped or the packet in consideration itself can be dropped depending on scheduling algorithm.  
	
	\item \textbf{Remove()}:  This interface's implementation picks the next $PHV$ to be transmitted from $SDS$ and decides when the packet will be actually transmitted through ${Port}_{egress}$. 
	\begin{gather*}
	f_{{S}^{remove}}: { {PHV\_SET} \, \rightarrow {PHVS}} \\
	f_{{S}^{remove}}(Null, Remove\: Interface \:Implemen\\
	tation, \, S_{param}) = PHV
	\end{gather*}
\end{itemize}

\textbf{CTP features}:  a) custom insert and remove interface: is it possible to provide implementation of these 2 interfaces for custom  scheduling algorithm b) custom SDS creation: custom SDS can be created for complex scheduling algorithm or not c) cross component access: can other components in the pipeline access SDS (ex. clear a queue based on certain event detected at match action unit) or it's properties (occupancy rate, priority etc.).

\textbf{RTC features}:
a) insert and remove interface parameter modifiability: can the control plane configure parameters for controlling behavior of  insert and remove functions b) SDS property access: can control plane  access SDS properties c) $S_{param}$ controlling: can CP control scheduling algorithm parameters at run time.

\subsection{Egress Port (${Port}_{E}$)} \label{EgressPort}
After a PHV is selected for emitting,  \textbf{\textit{egress port (${Port}_{E}$)}} transmits the content of $PHV.data\_buffer$. How a hardware level frame is created and transmitted is out of the scope of our discussion. Like ingress ports, egress ports also do not have any kind of programmability in terms of $CTP$ and $RTC$ features. 
\begin{gather*}
f_{{Port}_{In}}: {PHVS} \rightarrow {BS}  \\
f_{{Port}_{In}} (PHV,Null,Null) = PKT , \, 
\end{gather*}

\begin{table*}

	\caption{Few important algorithms and availability of relevant programmability features  }
	\label{tab:algoVsProduct}
	\begin{tabular*}{\textwidth}{l|l|l|l|l|l}
		\hline
		Application                                                                                     &  \begin{tabular}[c]{@{}c@{}}Required Programmability \\Features \end{tabular}                                         & \begin{tabular}[c]{@{}c@{}}Relevant AVS \\ Component\end{tabular}                         & FlexPipe & Arista 7170 & Agilio CX \\ \hline
		\begin{tabular}[c]{@{}c@{}}New protocol;\\ Telemetry;\end{tabular}                              & \begin{tabular}[c]{@{}c@{}}On the fly configuration \\ of packet parsing and \\ de-parsing logic\end{tabular} & Parser and deparser                                                                       & No       & Yes         & Yes       \\ \hline
		\begin{tabular}[c]{@{}c@{}}5g connectionless \\ communication;\\ Mobility management;\end{tabular} & \begin{tabular}[c]{@{}c@{}}Pausing packets at \\ intermediate switches \\ of a path\end{tabular}              & \begin{tabular}[c]{@{}c@{}}Ingress buffer engine\\ (programmable buffer)\end{tabular}     & No       & Partial     & Yes       \\ \hline
		\begin{tabular}[c]{@{}c@{}}Congestion Control;\\ Traffic engineering;\end{tabular}              & \begin{tabular}[c]{@{}c@{}}Programmable packet \\ scheduling\end{tabular}                                     & Scheduler                                                                                 & No       & No          & No        \\ \hline
		Video streaming;                                                                                & Manycast packet transfer                                                                                      & \begin{tabular}[c]{@{}c@{}}Buffer and replication \\ engine\end{tabular}                  & Partial  & Partial     & Yes       \\ \hline
		\begin{tabular}[c]{@{}c@{}}Traffic monitoring;\\ Flow tracking;\end{tabular}                    & \begin{tabular}[c]{@{}c@{}}Flow Identification;\\ Flow statistics;\end{tabular}                               & \begin{tabular}[c]{@{}c@{}}Match-action unit with\\ Stateful Data Structure;\end{tabular} & Partial  & Yes         & Yes       \\ \hline
	\end{tabular*}
\end{table*}

\section{Programmability Level Comparison} \label{ProgrammabilityFeatureMatrix}

PDP devices enables complex computation on packets in data plane. But limited memory, limited set of actions and requirement of maintaining line rate makes core switches unsuitable for complex algorithm. On the other hand,  smart-NIC based packet processing in slow path (cpu based processing) provides opportunity for more complex computation. But achieving high speed line rate in such environments still remains a big issue. To understand what kind of algorithms can be implemented in data plane with available  programmable switches in the current market, we have selected following representative platforms for comparison. \textbf{FlexPipe}\cite{flexpipeDatasheet} is Intel's Openflow supported programmable switching platform. It is selected as one of the early generation programmable switch of recent times. \textbf{Tofino}\cite{BarefootTofino}  is the most prominent and commercially successful programmable switching chip based on RMT architecture. It is being used by various switch vendors \cite{BarefootTofinoPartnerss}. It's various features are protected under barefoot non disclosure agreement. For comparison, we have chosen Arista 7170  \cite{arista7170} platform. It is based on Tofino chip and  supports P4\textsubscript{16}. \textbf{Netronome Agilio Cx} \cite{agilioCX}  (with  4000/6000 family NFP) is selected as a smnart-NIC based programmable switching platform.  It can be programmed using micro-C and P4.

In table \ref{tab:algoVsProduct}, we have listed few important types of algorithms, what are their crucial tasks that needs support from programmable switches (not achievable with traditional switches) and whether these tasks are supported by selected set of devices. From table \ref{tab:algoVsProduct}, it is clear that, all the devices in the market are not equally programmable. To better understand their programmability level, we have scored those selected platforms in table ~\ref{tab:funcationalProgrammabilityComparisonOfComponents} considering AVS components as base. Besides the selected programmable data plane platforms we have also included \textbf{PSA} \cite{psaSpec}. \textbf{PSA} is the most matured switch architecture by P4 community and it is based on P4\textsubscript{16}.  \textbf{\textit{AVS}} is also inspired by PSA. Until now, there is no hardware that follows full specification of PSA. But any architecture can be simulated in software level. Bmv2\cite{bmv2} is the reference implementation of P4, which  can simulate various switch architectures including PSA.  PSA is still not fully supported in bmv2. For comparison purpose, we assumed that PSA will support P4\textsubscript{16} and it will run on bmv2 simulation platform. Though PSA is still not realized, we believe comparison of existing hardware architecture with PSA (on bmv2) can give a clear picture of current state of the art. We have also included \textbf{T-switch} which represents most common category of traditional non-programmable L2/L3 switches for better understanding of the scores.

\begin{table}
	\caption{Eq. \ref{functionalDescriptionOfCompo} Based Programmability Comparison Matrix}
	\label{tab:funcationalProgrammabilityComparisonOfComponents}
	\begin{tabular}{@{}|c|c|c|c|c|c|@{}}
		\toprule
		\multicolumn{6}{|c|}{$E_1 = \mathcal{I}$, $E_2 = \mathcal{O}$, $E_3 = {Conf}_{param}$, $E_4 = {Proc}{Logic}$ }                                                                                                                                                                                                                                                                                             \\ \toprule
		{} & {} & \multicolumn{4}{|c|}{Features }
		\\ \toprule
		\multicolumn{1}{|c|}{Component} & Products & $E_1 $ & $E_2 $ & $E_3$ & $E_4$ \\ \midrule
		\multirow{4}{*}{\begin{tabular}[c]{@{}l@{}}\\ \\ \\Parser \end{tabular}}                                                                 & FlexPipe    & NA & 1        & NA         & 1 \\ \cmidrule (l){2-6} 
		& Arista 7170 & NA & 2        & NA         & 2 \\ \cmidrule (l){2-6} 
		& Agilio Cx   & NA & 2        & NA         & 2 \\ \cmidrule (l){2-6} 
		& 
		PSA        & NA & 2        & NA         & 2 \\ \cmidrule (l){2-6}
		& T-switch & NA & 0        & 0         & 0\\ \midrule
		\multirow{4}{*}{\begin{tabular}[c]{@{}l@{}}\\ Ingress \\Buffer \\ Engine\end{tabular}}                & FlexPipe    & 1 & 1                              & 1         & 0 \\ \cmidrule (l){2-6} 
		& Arista 7170 & 2 & 2                              & 1         & 0 \\ \cmidrule (l){2-6} 
		& Agilio Cx   & 2 & 2                              & 2         & 2 \\ \cmidrule (l){2-6} 
		& 
		PSA        & NA & 2                             & NA         & NA \\ \cmidrule (l){2-6}
		& T-switch & 0  & 0                             & 0         & 0\\ \midrule
		\multirow{4}{*}{\begin{tabular}[c]{@{}l@{}}\\ Ingress \\ Match\\Action\\ Unit\end{tabular}}            & FlexPipe    & 1 & 1                              & 1         & 1 \\ \cmidrule (l){2-6} 
		& Arista 7170 & 2 & 2                              & 2         & 3 \\ \cmidrule (l){2-6} 
		& Agilio Cx   & 2 & 2                              & 2         & 3 \\ \cmidrule (l){2-6} 
		& 
		PSA         & 2 & 2                              & 2         & 3 \\ \cmidrule (l){2-6}
		& T-switch & 0 & 0                              & 0         & 0\\ \midrule
		\multirow{4}{*}{\begin{tabular}[c]{@{}l@{}}\\ \\ \\Deparser\end{tabular}}                     & FlexPipe    & 1 & 2                              & NA         & 1 \\ \cmidrule (l){2-6} 
		& Arista 7170 & 2 & 2                              & NA         & 2 \\ \cmidrule (l){2-6} 
		& Agilio Cx   & 2 & 2                              & NA         & 2 \\ \cmidrule (l){2-6} 
		& 
		PSA         & 2 & 3                              & NA        & 2 \\ \cmidrule (l){2-6}           & 
		T-switch    & 0 & 3                              & NA        & 0\\ \midrule
		\multirow{4}{*}{\begin{tabular}[c]{@{}l@{}}\\Buffer \& \\ Replication \\Engine\\ (BRE)\end{tabular}} & FlexPipe    & 1 & 1                              & 1         & 0 \\ \cmidrule (l){2-6} 
		& Arista 7170 & 2 & 2                              & 1         & 0 \\ \cmidrule (l){2-6} 
		& Agilio Cx   & 2 & 2                              & 2         & 2 \\ \cmidrule (l){2-6} 
		& 
		PSA         & 2 & 2                              & 1         & 0 \\ \cmidrule (l){2-6}           & 
		T-switch    & 0 & 0                              & 0         & 0 \\ \midrule
		\multirow{4}{*}{\begin{tabular}[c]{@{}l@{}} \\ \\ Scheduler \end{tabular}}                                                                                     & FlexPipe    & 1 & 1                              & 1         & 1 \\ \cmidrule (l){2-6} 
		& Arista 7170 & 2 & 2                              & 1         & 1 \\ \cmidrule (l){2-6} 
		& Agilio Cx   & 2 & 2                              & 2         & 3 \\ \cmidrule (l){2-6} &
		PSA        & 2 & 2                              & 1         & 1 \\ \cmidrule (l){2-6}
		& T-switch & 0 & 0        & 0         & 0\\ \midrule
		
	\end{tabular}
\end{table}

Table ~\ref{tab:funcationalProgrammabilityComparisonOfComponents}, gives a 2-dimensional comparison among the selected platforms. Firstly, each component of AVS provides a specific type of programmablity in data plane. Comparing any platform on the basis of whether equivalent AVS components exists or not says
whether equivalent functionality can be achieved in a platform or not. Secondly, eqn. \ref{functionalDescriptionOfCompo} gives a functional structure  for each of the component. Scoring the platforms  based on parameters  ($\mathcal{I}$, $\mathcal{O}$, $ProcLogic$ and  ${Conf}_{param}$) of this equation gives a fine grained view of each platform's progammability level. Details comparison behind the scoring can be found in our full technical report \cite{AVSTechReport}.

The scoring system is  following: 
$\mathcal{I}, \mathcal{O}$ and ${Conf}_{param}$ represents some abstract data types. 
\begin{itemize}
\item \textbf{NA}- Programmability is not applicable here. Example: for all of the selected platforms, input  ($\mathcal{I}$) to ingress parser is a set of bits  ($PHV.data\_buffer$). Programmability is not needed for input to an ingress parser. 
\item \textbf{0}- The component is non programmable (ex. T-switch).
\item \textbf{1}- Selectable from a set of predefined data types. Example: FlexPipe \cite{flexpipeDatasheet}  hardware can only parse selected set of L2-L4 header fields.
\item \textbf{2}- New abstract data type  (i.e. struct. class etc.) can be created. Example: in P4 supported devices, packet header definition can be created from struct, enum etc.
\end{itemize}

FOR $ProcLogic$:
\begin{itemize}
\item \textbf{NA}- Programmability is not applicable here. Example: PSA has no provision for ingress buffer engine.
\item \textbf{0}- Non programmable component. Ex, traditional switch not supports any programmability for match-action units.
\item \textbf{1}- Only selectable from a set of pre-implemented logic /algorithm. Ex, in Flexpipe logic for match-action units are not programmable, only selectable from a predefined set of OpenFlow based actions.
\item \textbf{2}- New algorithms  ($ProcLogic$) can be implemented but stateful actions  (any action that can fetch and store a flow state) not supported. Ex. parser of  P4 supported switches can be programmed to parse headers but no stateful memory is not available here .
\item \textbf{3}- New algorithms  ($ProcLogic$) can be implemented with stateful actions support. Ex. P4 supported switches can perform action and store results in counter or register. 
\end{itemize}

\section{Motivating Use Cases} \label{MotivatingScenarios}
Abstraction plays very important role in SDN \cite{rfc7426}. Abstractions used in SDN are hierarchical in nature and used in different layers \cite{Casado:2014:ASN:2661061.2661063}. Hardware abstraction layer is placed over programmable hardware layer and provides a uniform view to other layers  (\textit{`Device and resource Abstraction Layer (DAL)'} \cite{rfc7426}). As an instance of  hardware abstraction layer, \textbf{\textit{AVS}} hides low level hardware complexity and it can provide a uniform view of hardware layer to both control plane and data plane application.  Hence its use cases can be found in every aspect of a truly programmable data plane device. We are mentioning few important use cases here. 

\subsection{Modular Architecture  Design \& Development} \label{ModularComponentDesignDevelopment}
Modular architecture design is the goal of good design. But majority of the programmable data plane architecture in literature are expressed in informal language. This makes modular design harder for 3 major reasons: a) lack of clear description about the role of a component b) lack of clear boundary between 2 components c) how the components connect with each other in pipeline. These shortcomings create bottleneck in independent design and  optimization of  components. It also brings disadvantages in designing new pipeline based on reuse of those components.  A modular abstraction layer can solve these issues.

Consider a PDP architecture, where boundary between \textbf{\textit{Ingress Buffer Engine  (${BE}_{In}$)}} and \textbf{\textit{Ingress Match Action Unit (\textbf{\textit{${MAU}_{In}$}})}} is not clear . As both the components have match-action semantics, ${BE}_{In}$ can be designed as a part of ${MAU}_{In}$. After header matching, storing in and removing packet from buffer can be designed as one of the actions of ${MAU}_{In}$. Now, consider a smart-NIC based packet processing architecture, where packets are moved to userspace and processed in CPU and/or GPU. Also assume ${MAU}_{In}$ is implemented on CPU and GPU . GPUs perform better in parallel and batch processing, where buffering is a per packet processing task not best suitable for GPU.  Implementing buffer engine as part of match-action unit using GPU needs costly data transfer to and from memory to GPU. Or for smart-NIC based environments packets are needed to be moved from smart-NIC buffer to GPU. This data movement often loses the GPU performance gain. Hence it is better to implement buffer engine as a separate component. If buffer engine functionality is merged with ${MAU}_{In}$ optimal hardware performance can not be achieved in this case. 
In this example, if \textbf{\textit{AVS}} like abstraction layer is used, data plane application developers can write logic without thinking about underlying CPU and/or GPU based implementation. \textbf{\textit{AVS}} components to actual hardware mapping is done by the compilers. Thus tasks of ${BE}_{In}$ can be executed on CPU only and ${MAU}_{In}$ can be mapped to CPU and/or GPU depending on performance goal. Moreover, new hardware (ex. specialized chip or FPGA) can be designed and optimized independently for each of the components. 

\subsection{Virtualization \& New Pipeline Design} \label{VirtualizationNewPipelineDesign}

Equation \ref{functionalDescriptionOfCompo} provides a structured framework for each component. It 
clearly defines the input, output, packet processing logic and parameters for how to configure run-time behavior of a component through control plane  (${Conf}_{param}$). Communication among the components is only  through passing parameters.  It removes control dependency among components. This gives several advantages:  a) AVS gives a hardware independent and portable representation of data plane device which can be used to slicing the hardware layer b) DPPs can be developed based on machine model provided by \textbf{\textit{AVS}} c) \textbf{\textit{AVS}} can be used to create a virtual switch (VM like entity) for data plane d) switch hypervisor like entities can execute or migrate same DPP to different hardware architecture.

Consider a smart-NIC based scenario where the datapath is moved to userspace using DPDK. Also assume, the data plane prorgam  (\textbf{DPP}) is assigned to do both  IPv4/6 packet processing and running an algorithm for accelerating ML algorithms through in network aggregation technique \cite{sapio2019scaling,wang2019high}. Aggregation is basically mathematical processing and computationally expensive which can be executed in parallel fashion. Let, IPv4/6 packet processing is expressed as ${MAT}_{IP}$ and aggregation based packet processing logic is expressed as ${MAT}_{AGG}$. Now, to make packet aggregation faster, ${MAT}_{AGG}$ is assigned to be executed on GPU. And IPv4/6 packet processing  ( ${MAT}_{IP}$) tasks are executed on CPU. Also assume, CPU can process 1 IPv4/6 packet in 1 cycle where as GPU can aggregate 10 packets in a single cycle and produce result of aggregation as one packet. As CPU and GPU run in different speed, synchronizing and scheduling the packet processing over them is very important. Without a clear definition and structure of  ${MAT}_{IP}$ and  ${MAT}_{AGG}$,  synchronizing packet processing over CPU and GPU is not possible. Moreover based on underlying server capability number of CPU cores may differ and GPUs may not be available at certain time. To handle these kind of scenarios, a VM like entity for data plane is mandatory. \textbf{\textit{AVS}} can works as a virtual switch for data plane.Moreover, modular representation provided by \textbf{\textit{AVS}}, can also enable switch hypervisors to assign and synchronize execution of DPP over various types of hardware.

\subsection{Testing \& Verification} \label{TestingAndVerification}
Earlier data plane devices were mainly designed for executing network protocol. But in programmable data plane devices, more and more application layers tasks are pushed toward data plane. These devices can be loaded with new data plane program at any period of their life time. Testing them or validating some property at run time over these programs are very important. Without a common abstraction layer and workflow, testing programs and verifying properties over heterogeneous hardware architecture increases the cost and complexity. To write test cases, a common set of packet processing state among switches are required. Without a common abstraction layer and a common set of packet processing states, how various hardware vendors  implements packet processing logic can differ.

\begin{figure}
	\centering
	\includegraphics[ width=\columnwidth, trim={0.0in 5.8in 6.72in 0.0in},clip]{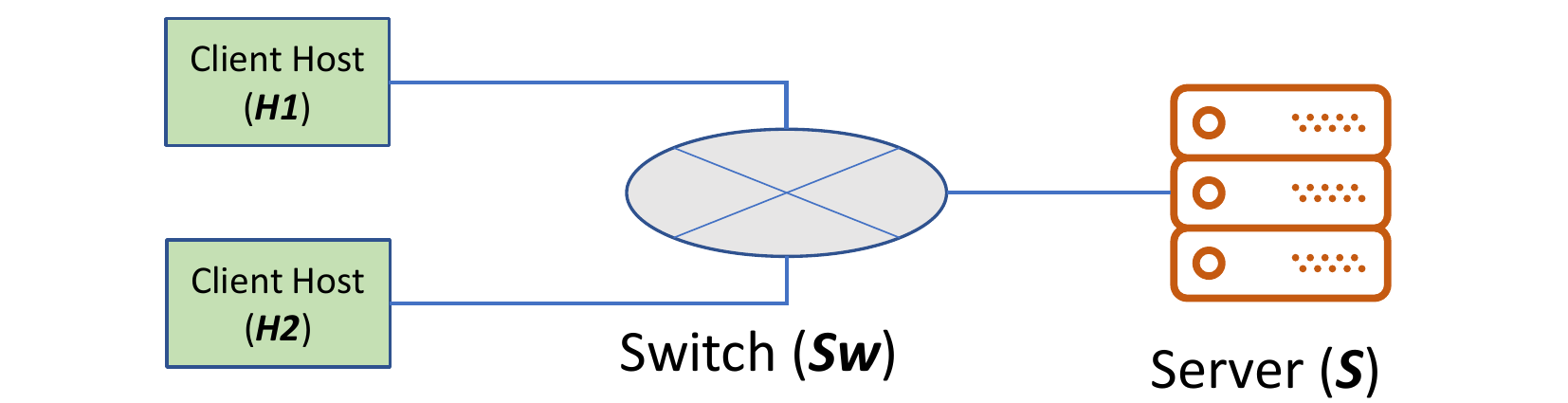}
	\caption{Server-client communication example}
	\label{fig:scenario3}
\end{figure}

Consider client-server  scenario of Fig. ~\ref{fig:scenario3}. At a certain time \textbf{\textit{H1}} got disconnected from server (\textbf{\textit{S}}) due to  \textbf{\textit{H1-Sw}} link failure. Detecting the link failure switch (\textbf{\textit{Sw}}) stores all packets directed for \textbf{\textit{H1}} in \textbf{\textit{$B_i$}} of ingress buffer engine (section \ref{bufferSubSection}) . When \textbf{\textit{$B_i$}} becomes full, \textbf{\textit{Sw}} sends a notification to controller.  Upon receiving this notification, controller initiates migration of the whole network  (migration including switch state a packet stored in buffer \cite{Keller:2012:LME:2390231.2390250}) from one data center to another. Clearly, this kind of DPP depends on the \textbf{\textit{"$B_i$ is Full"}} state of buffer receiver thread (sec. \ref{bufferSubSection}, Fig. \ref{fig:bufferRcvr}). For testing them and ensuring uniform behavior, switch vendor of both the data center need to support ingress buffer engine with common packet processing states. Use of \textbf{\textit{AVS}} as an abstraction layer provides a common abstract switch over heterogeneous architecture and  DPP developer gets a uniform view of the hardware. Moreover \textbf{\textit{AVS}} expresses behavior of the components through EFSM. Common abstraction of the components and common set of packet processing states of the components can be leveraged together to write portable code and testing them.

\subsection{Network Function Modeling} \label{NetworkFunctionModeling}   
Eq. \eqref{functionalDescriptionOfCompo} enables representation of  each components of \textbf{\textit{AVS}} in a uniform and hardware independent manner. Composing $f_c$ for all the components and concatenating them represents \textbf{\textit{AVS}} as a transfer function ($\tau$) \citep{180587}.  Compiler translates this functional representation to hardware instruction. Any network function (NF) programmed on \textbf{\textit{AVS}} can be expressed as $\tau$, where $\mathcal{I}$, $\mathcal{O}$, $Conf_{Param}$, $Proc_{Logic}$ are  different for each NF . Each link of a network transfers a $PKT$ from one hop to another. Link also can be represented as a function ($\Gamma$) of same structure.
\begin{gather*} \label{functionalDescriptionOfLink}
{\Gamma}: \mathcal{BS} \rightarrow \mathcal{BS}  \\
\Gamma (PKT,Medium \, Access \, Logic,{Conf}_{param}) = PKT
\end{gather*} 
$\Gamma$ can be considered as the functional representation of data link layer.  Depending on medium, corresponding $Medium$ \\ $Access \, Logic,$ (i.e. ring/mesh/star networks, CSMA, CSMA/ CD etc.) and  ${Conf}_{param}$ (i.e. persistent level for CSMA) can be different. Based on these, a $PKT$'s transmission and propagation delay can be different. These delays can be derived from time stamp (passed as $H_f$) at sender node and receiver node. Again applying $\tau$ and $\Gamma$ a whole network topology can be represented as a topology function ($\psi$) \cite{180587}. These functional representation  are based on abstraction layer and they are hardware independent representation with  well defined parameters for each component. This kind of representation provides various benefits. Few of them are 

\begin{itemize}
\item \textbf{Network algebra}: Network algebra is an old and deeply studied topics. With rise of programmable data plane device, it is more relevant and applicable for today's network. Expressing  network functions and networks in a functional paradigm allows usage of formal methods of network algebra  \cite{Anderson:2014:NSF:2578855.2535862} \cite{180587}.

\item \textbf{Network Delay Modeling}: Time-stamping \cite{7562197} packet in data plane is a strong concept with many usage \cite{mymb-sfc-nsh-allocation-timestamp-05}. But maintaining time-stamp only at entry and exit point of a switch can't provide deep information about various type of delay. Time-stamp  can be maintained at entry (${time\mbox{-}stamp}_{C}^{Entry}$) and exit (${time\mbox{-}stamp}_{C}^{Exit}$) point of   each component in packet metadata (${PKT}_{Metadata}$). Delay inside a component ($\mathcal{C}$) can be calculated as 
\begin{equation*}
{D}_{\mathcal{C}} =  {time\mbox{-}stamp}_{C}^{Exit} \, -  \, {time\mbox{-}stamp}_{C}^{Entry} 
\end{equation*}
Thus, fine grained information about various delay can be collected. One of the most simple but powerful model of delay is Network Delay. It can be computed as following
\begin{gather*}
Network \, Delay \, = \\
					Queuing \, Delay \, + \, Processing \, Delay \,  + \\
					 (Transmission \, Delay\, + \, Propagation \, Delay) \\
					= \, {D}_{{BE}_{In}} \, + \, ( {D}_{{PR}_{In}} \, + \, {D}_{{MAU}_{In}} \, + \, 								{D}_{{DPR}_{In}} \\             
					+ \, {D}_{{BRE}} \, + \, {D}_{{PR}_{E}} \, + \, {D}_{{MAU}_{E}} \, + \, 
					{D}_{{DPR}_{E}} \\ 
					+ \, {D}_{S} )\, + \, {D}_{{PR}_{In}} \, + \, {D}_{\Gamma}
\end{gather*}

Complex data plane processing with time stamp can be used for further complex  measurement of various types of delay  categorized by source \cite{6967689}.
\end{itemize}

\section{Conclusion} \label{Conclusion}
In this work, we proposed the design of \textbf{\textit{AVS}}, a modular hardware abstraction layer for programmable data plane devices.  We are working on a bmv2 \cite{bmv2} based implementation of \textbf{\textit{AVS}}. Our work on representing  components of abstraction layer for data plane devices with a well defined functional structure and their work flow  in EFSM  can provide a strong base for optimizing, bench marking and comparing different programmable data plane devices. We invite research community to investigate how to improve this formal structure and develop frameworks for using it. We have also analyzed the programmability features of different components and compared few important products based on them. Deriving formal relation between what set of programmability features can support a specific class of algorithms in data plane can be a promising research direction. Besides this, designing chip directly from abstract model also can be a very promising research scope.


%
%

\bibliographystyle{acm}      
\bibliography{AVS}

\end{document}